\documentclass[twocolumn,trackchanges,twocolappendix]{aastex701}

\usepackage{longtable}
\usepackage{booktabs}
\usepackage{caption}
\begin{document}

\title{The JWST Sub-Jupiters Survey: The impact of a large JWST/NIRCam PSF library on exoplanet imaging detection sensitivity}

\correspondingauthor{Rodrigo Ferrer-Chavez}
\email{rodrigoferrer-chavez2029@u.northwestern.edu}

\author[0000-0002-2683-2396]{Rodrigo Ferrer-Chavez}
\affiliation{Department of Physics and Astronomy, Northwestern University, 2145 Sheridan Rd, Evanston, IL 60208, USA}
\affiliation{Center for Interdisciplinary Exploration and Research in Astrophysics (CIERA), Northwestern University, 1800 Sherman Ave, Evanston, IL 60201, USA}
\email{rodrigoferrer-chavez2029@u.northwestern.edu}

\author[0000-0003-0774-6502]{Jason Wang}
\affiliation{Department of Physics and Astronomy, Northwestern University, 2145 Sheridan Rd, Evanston, IL 60208, USA}
\affiliation{Center for Interdisciplinary Exploration and Research in Astrophysics (CIERA), Northwestern University, 1800 Sherman Ave, Evanston, IL 60201, USA}
\email{jason.wang@northwestern.edu}

\author[0000-0001-5365-4815]{Aarynn L. Carter}
\affiliation{Space Telescope Science Institute (STScI), 3700 San Martin Drive, Baltimore, MD 21218, USA}
\email{aacarter@stsci.edu}

\author[0000-0003-4614-7035]{Beth Biller}
\affiliation{Scottish Universities Physics Alliance, Institute for Astronomy, University of Edinburgh, Blackford Hill, Edinburgh EH9 3HJ, UK} 
\affiliation{Centre for Exoplanet Science, University of Edinburgh, Edinburgh EH9 3HJ, UK}
\email{bbiller@ed.ac.uk}

\author[0000-0002-7325-5990]{Ellis Bogat}
\affiliation{Department of Astronomy, University of Maryland, College Park, 4296 Stadium Dr, College Park, MD 20742, USA}
\email{Ebogat@umd.edu}

\author[0009-0009-1074-3696]{Rohan Kane}
\affiliation{Space Telescope Science Institute (STScI), 3700 San Martin Drive, Baltimore, MD 21218, USA} 
\email{rohan.kane22@gmail.com}

\author[0000-0002-1652-420X]{Giovanni M. Strampelli}
\affiliation{Institute for Astronomy, University of Edinburgh, Royal Observatory, Blackford Hill, Edinburgh, EH9 3HJ, UK}
\email{gstrampelli@roe.ac.uk}

\author[0000-0001-6396-8439]{William O. Balmer}
\altaffiliation{51 Pegasi b Fellow}
\affiliation{Center for Interdisciplinary Exploration and Research in Astrophysics (CIERA), Northwestern University, 1800 Sherman Ave, Evanston, IL 60201, USA}
\email{balmer@northwestern.edu}

\author[0000-0001-8568-6336]{Mark Booth}
\affiliation{UK Astronomy Technology Centre, Royal Observatory Edinburgh, Blackford Hill, Edinburgh EH9 3HJ, UK}
\email{markbooth@cantab.net}

\author[0000-0002-3191-8151]{Marshall Perrin}
\affiliation{Space Telescope Science Institute (STScI), 3700 San Martin Drive, Baltimore, MD 21218, USA}
\email{mperrin@stsci.edu}

\author[0000-0002-1838-4757]{Aniket Sanghi}
\altaffiliation{NSF Graduate Research Fellow}
\affiliation{Cahill Center for Astronomy and Astrophysics, California Institute of Technology, 1200 E. California Boulevard, MC 249-17, Pasadena, CA 91125, USA}
\email{asanghi@caltech.edu}

\author[0009-0000-0303-2145]{Rapha\"el Bendahan-West}
\affiliation{Department of Physics and Astronomy, University of Exeter, Stocker Road, Exeter EX4 4QL, UK}
\email{rb941@exeter.ac.uk}

\author[0009-0001-0275-7811]{Evelyn L. Bruinsma}
\affiliation{Department of Physics \& Astronomy, Johns Hopkins University, 3400 N. Charles Street, Baltimore, MD 21218, USA}
\email{ebruins1@jhu.edu}

\author[0000-0002-9962-132X]{Ben J. Sutlieff}
\affiliation{Institute for Astronomy, University of Edinburgh, Royal Observatory, Blackford Hill, Edinburgh, EH9 3HJ, UK} 
\affiliation{Centre for Exoplanet Science, University of Edinburgh, Edinburgh EH9 3HJ, UK}
\email{ben.sutlieff@roe.ac.uk}

\author[0009-0005-6943-6819]{Andrew D. James}
\affiliation{Department of Physics and Astronomy, University of Exeter, Stocker Road, Exeter EX4 4QL, UK}
\email{adj207@exeter.ac.uk}

\author[0009-0008-2252-7969]{Klaus Subbotina Stephenson}
\affiliation{Department of Physics and Astronomy, University of Victoria, 3800 Finnerty Road, Victoria, BC, V8P 5C2, Canada}
\affiliation{ NRC Herzberg Astronomy and Astrophysics Research Centre 5071 W Saanich Rd, Victoria, BC, V9E 2E7, Canada}
\email{klausss@uvic.ca}

\author[0000-0003-4909-256X]{Katie A. Crotts}
\affiliation{Space Telescope Science Institute (STScI), 3700 San Martin Drive, Baltimore, MD 21218, USA}
\email{kcrotts@stsci.edu}

\author[0000-0001-6098-3924]{Andy Skemer}
\affiliation{Department of Astronomy \& Astrophysics, University of California, Santa Cruz, 1156 High Street, Santa Cruz, CA 95064}
\email{askemer@ucsc.edu}

\author[0000-0001-5831-9530]{Rachel Bowens-Rubin}
\affiliation{Eureka Scientific Inc., 2542 Delmar Ave., Suite 100, Oakland, CA 94602, USA}
\email{rbowens-rubin@mit.edu}

\author[0000-0002-4388-6417]{Isabel Rebollido}
\affiliation{European Space Agency (ESA), European Space Astronomy Centre (ESAC), Camino Bajo del Castillo s/n, 28692 Villanueva de la Ca\~nada, Madrid, Spain}
\email{isabel.rebollidovazquez@esa.int}

\author[0000-0003-2769-0438]{Jens Kammerer}
\affiliation{European Southern Observatory, Karl-Schwarzschild-Straße 2, 85748 Garching, Germany}
\email{jkammere@eso.org}

\author[0000-0002-5352-2924]{Sebastian Marino}
\affiliation{Department of Physics and Astronomy, University of Exeter, Stocker Road, Exeter EX4 4QL, UK}
\email{s.marino-estay@exeter.ac.uk}

\author[0000-0001-8627-0404]{Julien H. Girard}
\affiliation{Space Telescope Science Institute (STScI), 3700 San Martin Drive, Baltimore, MD 21218, USA}
\email{jgirard@stsci.edu}

\author[0000-0002-6964-8732]{Kellen Lawson}
\affiliation{Center for Space Sciences and Technology, University of Maryland, Baltimore County, 1000 Hilltop Circle, Baltimore, MD 21250, USA}
\affiliation{Astrophysics Science Division, NASA-GSFC, 8800 Greenbelt Rd, Greenbelt, MD 20771, USA}
\affiliation{Center for Research and Exploration in Space Science and Technology, NASA-GSFC, 8800 Greenbelt Rd, Greenbelt, MD 20771, USA}
\email{kellen.d.lawson@nasa.gov}

\author[0000-0003-4203-9715]{Emily Rickman}
\affiliation{European Space Agency (ESA), ESA Office, Space Telescope Science Institute (STScI), 3700 San Martin Drive, Baltimore, MD 21218, USA}
\email{erickman@stsci.edu}

\author[0000-0002-0834-6140]{Jarron M. Leisenring}
\affiliation{Department of Astronomy / Steward Observatory, University of Arizona, Tucson, AZ 85721, USA}
\email{jarronl@arizona.edu}

\author[0000-0002-2428-9932]{Clémence Fontanive}
\affiliation{Institute for Astronomy, University of Edinburgh, Blackford Hill, Edinburgh EH9 3HJ, UK}
\email{clemence.fontanive@roe.ac.uk}

\author[0000-0002-9372-5056]{Nour Skaf}
\affiliation{Institute for Astronomy, University of Hawai'i, 2680 Woodlawn Drive, Honolulu HI 96822}
\email{nskaf@hawaii.edu}

\author[0000-0001-6205-9233]{Maxwell A. Millar-Blanchaer}
\affiliation{Department of Physics, University of California, Santa Barbara, CA 93106, USA}
\email{maxmb@ucsb.edu}

\author[0000-0001-8074-2562]{Sasha Hinkley}
\affiliation{Department of Physics and Astronomy, University of Exeter, Stocker Road, Exeter EX4 4QL, UK}
\email{S.Hinkley@exeter.ac.uk}

\author[0009-0004-0868-1186]{Aiza Kenzhebekova}
\affiliation{Institute for Astronomy, University of Edinburgh, Royal Observatory, Blackford Hill, Edinburgh, EH9 3HJ, UK}
\email{a.kenzhebekova@ed.ac.uk}

\author[0009-0006-3044-9011]{Mia Belle Parkinson}
\affiliation{UK Centre for Astrobiology, School of Physics and Astronomy, University of Edinburgh, King's Buildings, Edinburgh EH9 3FD}
\email{m.b.parkinson@sms.ed.ac.uk}

\author{Edward L. Turner}
\affiliation{Scottish Universities Physics Alliance, Institute for Astronomy, University of Edinburgh, Royal Observatory, Blackford Hill, Edinburgh, EH9 3HJ, UK}
\email{E.Turner-11@sms.ed.ac.uk}

\begin{abstract}

The excellent infrared sensitivity of JWST allows direct imaging of young low-mass (below $\sim1$ M$_{\text{Jup}}$) giant planets for the first time, with several JWST programs being carried out for this purpose. However, the detection sensitivity of these observations crucially depends on subtracting the residual light of the host star to reveal faint astrophysical signals. This requires high quality observations of reference stars to build an empirical starlight model. Here we present a homogeneously processed reference star library of targets observed with programs GO 4050, GO 5835 and SURVEY 6005, which enable PSF subtraction via reference star differential imaging. These observations were taken with the \texttt{MASK335R} coronagraph mask and in dual-filter F444W/F200W mode. Our library consists of 247 observations (counting both filters separately) of 100 different stars of spectral types mostly F to M, with a total of 1110 individual integrations. We explore the PSF subtraction hyper-parameters with the KLIP algorithm \citep{Soummer2012} that maximize sensitivity, and we show the performance of our library compared to and in addition to contemporaneous reference observations. We show that our library improves sensitivity at close separations in most tested cases, with improvement factors from $\sim1.3$ to $\sim2$ at some separations. As practical recommendations, we find that our library is best used as a complement to contemporaneous reference observations to further improve sensitivity, since in all our experiments contemporaneous references are needed for best sensitivity.  We make the PSF library publicly available for the high-contrast imaging community.

\end{abstract}



\section{Introduction}
\label{s:intro}

With its exceptional IR sensitivity, JWST is capable of directly imaging much lower mass and much colder giant planets than any other current observatory. This opens previously unexplored regions of exoplanet parameter space. On exoplanet demographics, constraints on the occurrence rate of wide-separation sub-Jupiters become possible for the first time \citep{carter2021direct, Carter_2023}. On atmospheric studies, imaging of planets at record-low masses and temperatures can yield insights into planet formation and composition (e.g., \citealt{bardalez2025jwst,matthews2024temperate, lagrange2025evidence,sanghi2026worlds}).

However, the sensitivity of high-contrast coronagraphic imaging observations is not only limited by the intrinsic sensitivity of JWST, but by our ability to remove residual starlight from our observations. Diffracted starlight in the stellar point spread function (PSF) has a complex morphology, often with patterns that look identical to planet signals \citep{racine1999speckle, hinkley2007temporal, martinez2013speckle}. Especially at close angular separations from the star (e.g. within $\sim3$ arcseconds for JWST/NIRCam in the F444W filter), it is our ability to subtract starlight without over-subtracting exoplanet light that sets the imaging sensitivity floor \citep{perrin2018updated}. This is particularly important for JWST, since it has lower angular resolution than 8-10 meter ground-based telescopes\footnote{For reference, the pixel scale of the long wavelength channel of JWST/NIRCam is $\sim 63$ mas \citep{Kammerer_2022}, compared to $\sim14$ mas for the Gemini Planet Imager \citep{macintosh2014first} and $\sim12$ mas for SPHERE/IRDIS \citep{maire2016sphere}}. This means even long-period planets lie at very small angular separations from the star (for JWST/NIRCam in the F444W filter, a separation of 5 au around a very nearby star at 10 pc corresponds to only 8 pixels). Consequently, planets must routinely be searched for or recovered right in the regime where PSF subtraction is most challenging.


In ground-based observatories, subtraction of the stellar PSF is mainly done through angular differential imaging (ADI; \citealt{liu2004substructure,marois2006angular}). In ADI, the planet rotates azimuthally around the star due to the rotation of the Earth. This allows us to disentangle the spatially static signal of the stellar PSF from the rotating planet signal. A large set of observations at different parallactic angles are typically combined through principal component analysis (PCA)-based algorithms (e.g., KLIP, \citealt{Soummer2012}). These construct empirical PSF models from the principal components of the set of observations. 

With JWST, the utility of ADI-only PSF subtraction is limited because of the restricted amount of rotation available to JWST ($\sim10^\circ$). Instead, PSF subtraction is typically done via reference star differential imaging (RDI). In RDI, a bright reference star is observed right before or after a science sequence. Algorithms like KLIP can then be used to construct a PSF model from the reference observations and subtract it from the science image. While this strategy is effective (e.g., \citealt{smith1984circumstellar, lagrange2009probable, lafreniere2009hst}), it is known that the PSF of the reference star and the PSF of the science target are never identical. This is for several reasons: both stars can have different spectral energy distributions, aberrations in the telescope are constantly evolving, and imperfect pointing accuracy. As a result, this strategy does not guarantee an ideal PSF subtraction.

Additionally, acquiring reference star observations for every science target is an expensive use of telescope time, especially given that the reference star is typically not of scientific interest. For this reason, several JWST programs have opted for a ``self-referencing" strategy, where a handful of science targets are observed back-to-back in sequence. In this way, the science targets themselves can be used as reference star PSFs for other science targets. While this represents a gain in efficiency, this strategy comes at the obvious cost of less-than-ideal reference PSFs, and thus (at least \textit{prima facie}) a sacrifice in per-observation sensitivity.

A clear way to enhance the sensitivity of high-contrast observations with JWST is a large reference PSF library. A large set of observed PSFs is likely to have enough diversity to capture most details in most PSFs, enabling effective PSF removal. Furthermore, even when a dedicated reference star is available, it is likely that adding a large PSF library may further improve sensitivity (e.g., \citealt{girard2022}). The use of large  PSF libraries has precedents both in ground-based observatories (e.g., \citealt{xie2022reference, sanghi2024efficiently}) and in space with the Hubble Space Telescope (e.g., \citealt{soummer2014five, choquet2016first, Ren2017,sanghi2022efficiently}).

Here we present a compilation of a large PSF library using data from various JWST programs collectively referred to as \textit{the JWST Sub-Jupiters Survey}: GO 4050, GO 5835 and SURVEY 6005. All of these programs share the same observing setup and thus can serve as suitable reference PSFs for each other. These are also common observing setups for exoplanet imaging in general, so this library can be used to improve sensitivity (particularly at close angular separations of the star) in a large number of datasets. The library presented here was used for the discovery of giant planet RX J0534 b \citep{ferrer2026jwst}, and also, partially or completely, for other analyses of JWST/NIRCam observations \citep{palatnick2025,crotts2026jwst}.

This paper is organized as follows: in Section \ref{s:obs_desc} we describe the surveys our library was constructed from; Section \ref{s:construct} outlines the data reduction and the PSF library construction; Section \ref{s:contrast_opt} describes a suite of tests performed to identify PSF subtraction parameters that maximize sensitivity with our library; Section \ref{s:results_grids} describes the best sensitivities achieved in the background-limited regime and the speckle-limited regime; in Section \ref{s:bigcomparison} we investigate the sensitivity improvements of using a large PSF library. Finally, we discuss some of our findings in Section \ref{s:discussion}, including practical recommendations for JWST post-processing, and conclude in Section \ref{s:conclusions}.


\section{Observation description}
\label{s:obs_desc}

The PSF library presented here includes observations from three separate JWST programs that share both science goals and observational setups, such that targets observed in any program can in principle be used as reference stars for any other. The three programs are:

\begin{itemize}
    \item JWST GO 4050 (``Uncharted Worlds: Towards a Legacy of Direct Imaging of Sub-Jupiter Mass Exoplanets"; PI: A. Carter) targeted a sample of young ($\sim10$ Myr) M-stars in the nearby ($\sim50$ pc) TWA association. The sample consists of 23 science targets observed with no dithering and 7 dedicated reference stars observed with the \texttt{9-POINT-CIRCLE} small grid dither pattern.
    \item JWST GO 5835 (``Into The Spotlight: Unveiling Wide-Separation Sub-Jupiters for Future JWST Characterization"; PI: A. Carter) targeted a sample of young ($\sim26$ Myr) M-stars in the nearby ($\sim30$ pc) $\beta$ Pictoris moving group. The sample consists of 45 science targets observed with no dithering and no dedicated reference stars with dither patterns. Additionally, this program has 11 targets observed with non-coronagraphic direct imaging which are not included in this work.
    \item JWST SURVEY 6005 (``Imaging Young Sub-Jupiter Planets down to Solar-System Scales"; PI: B. Biller) targeted a sample of young ($<500$ Myr, and mostly $<150$ Myr) FGK-stars within 60 pc. The sample consists of 96 science targets observed with no dithering and no dedicated reference stars with dither patterns. Due to significant errors in coronagraph alignment, we exclude 10 of these targets, bringing the sample to 86 science targets.
\end{itemize}

Future versions of the PSF library may include archival observations with the same observational setup for increased PSF diversity.

All observations were taken with the NIRCam coronagraph instrument \citep{girard2022} using the \texttt{MASK335R} mask, in the \texttt{SUB320A335R} sub-array, and in dual-filter mode for simultaneous observations with the F444W and F200W filters. The F444W filter yields the highest sensitivity to low-mass giant planet detections \citep{Carter_2023}. The F200W filter enables F200W$-$F444W color calculations and therefore a preliminary distinction between bonafide planet candidates and background galaxies or stars (e.g., \citealt{Crotts2025, palatnick2025, bardalez2025jwst}). The field of view (FOV) for the F444W data is 20$\arcsec$ $\times$20$\arcsec$ (pixel scale: $\sim$ 63 mas per pixel), and for the F200W data it is 10$\arcsec$ $\times$ 10$\arcsec$ (pixel scale: $\sim$ 31 mas per pixel). All targets were observed without position angle rotation to maximize survey efficiency.

Below we describe the construction of our PSF library from the data of these surveys. Given that all the observations analyzed here use the JWST/NIRCam  MASK335R coronagraph, throughout the manuscript we refer to the stellar coronagraphic PSF simply as \textit{PSF} or \textit{stellar PSF} for brevity, except when specified otherwise.

\begin{figure*}
    \centering
    \includegraphics[width=1\linewidth]{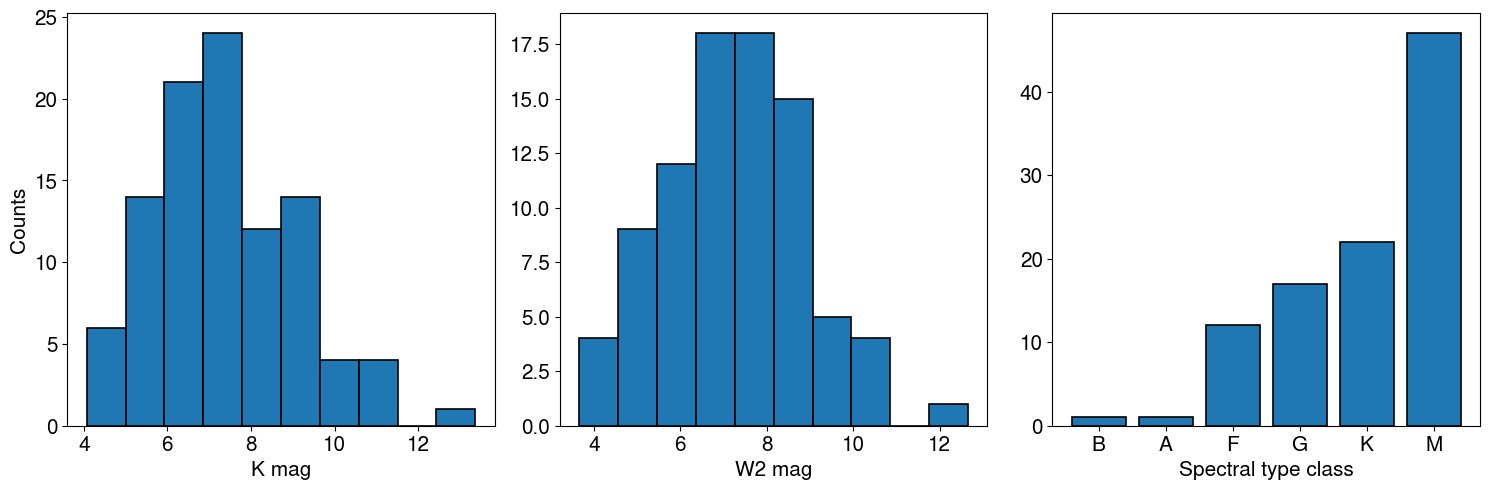}
    \caption{Properties of stars included in the PSF library. Programs GO 4050 and GO 5835 target predominantly M stars (in the TW Hya and $\beta$ Pictoris associations, respectively), and SURVEY 6005 targets a variety of stellar associations within $\sim60$ pc, mostly FGK stars. See Table \ref{tab:properties} for the values and references used.}
        \label{fig:histograms}
\end{figure*}
\section{PSF library construction}
\label{s:construct}

\subsection{Data pre-processing}
\label{s:preprocessing}

All observations were processed with \texttt{spaceKLIP} \footnote{\url{https://github.com/spacetelescope/spaceKLIP}} \citep{Kammerer_2022, Carter_2023}, a publicly available code that integrates the standard \texttt{jwst} pipeline with custom data reduction steps for high-contrast imaging. \texttt{spaceKLIP} also includes tools for PSF subtraction and companion extraction with the KLIP \citep{Soummer2012} and KLIP-FM \citep{Pueyo_2016} algorithms implemented in \texttt{pyKLIP} \citep{Wang_2015}.

The raw \texttt{uncal.fits} files were downloaded from the Mikulski Archive for Space Telescopes (\texttt{MAST}) and processed uniformly with the standard \texttt{spaceKLIP} pipeline. We used the \texttt{jwst} pipeline version 1.20.2 and \texttt{spaceKLIP} version 2.2.1 (GitHub commit hash: 6be81e4d). 

All observations are uniformly reduced using stages 1 and 2 of the standard JWST pipeline\footnote{\url{https://jwst-pipeline.readthedocs.io/en/latest/jwst/pipeline/calwebb_detector1.html}}$^,$\footnote{\url{https://jwst-pipeline.readthedocs.io/en/latest/jwst/pipeline/calwebb_image2.html}}. Through \texttt{spaceKLIP} we apply additional data preparation steps for improved PSF subtraction, including background subtraction and a custom bad pixel correction routine. We do not stack or co-add the various integrations or dithers for each target, so during PSF subtraction each integration is treated as a separate image. See Appendix for details on the data reduction.

\subsection{Target selection for the library}
\label{s:selection}

After uniformly processing all the observations from all three programs, we selected the subset of them that were suitable to be included as reference PSFs. 

First, we identified several observations with overly bright off-axis sources, with astrophysical contaminants very close to the inner PSF (within $\sim3$ arcsec for the F444W filter and $\sim1.5$ arcsec for F200W) or with fields of view crowded with a large number of faint background contaminants or cosmic rays. These are contaminants with non-negligible signal in large portions of the image, or that contaminate the region of most interest (close angular separations), such that they are difficult to properly mask. Therefore, we excluded these from the PSF library to avoid the accidental injection of contaminants during PSF subtraction. We additionally excluded two observations with bad integrations (possibly affected by tilt events, \cite{rigby2023}, or other instrumental instability) since they exhibited radically different PSFs from the rest of the targets in the survey.

Next, we carried out PSF subtractions for each of the targets in the library using all other observations in the library as reference stars, using all available KL modes and one single annulus in all subtractions. These initial subtractions allow us to identify additional contaminants hiding under the stellar PSF, such as disks, stars\footnote{All visual binaries were identified by eye and a more in-depth investigation on the binaries is planned. More details are discussed in Strampelli et al (in prep).}, galaxies or sub-stellar objects within $\sim3$ and $\sim1.5$ arcsec of the star for the F444W and F200W filters, respectively. Given the spatially correlated noise after PSF subtraction and the diversity in morphology of possible contaminants, it is difficult to identify them with an automated pipeline (e.g., using match filtering). Therefore, we opted for a collaboration-wide effort to identify these contaminants via manual inspection. After manually flagging observations with contaminants at close angular separations, we exclude these from our PSF library. From a total of 444 observations (counting both filters separately), we end up with 247 in our library ($\sim56\%$ of the total). 

The remaining observations might still have a few contaminants in the field of view. By construction, these are faint enough and far enough from the stellar PSF that it is possible to clean them without negatively affecting PSF subtraction. To do this, we take the location of every one of these contaminant sources (found from the same process described above) and replaced the pixels within a circular mask centered at the source with a radius sufficient to block all the pixels affected. To replace the masked region while preserving PSF structure, we produced several ``template" PSFs by computing the median value per pixel across sets of contaminant-free observations (see below). We replace the pixels in the masked regions with the stellar flux-weighted values of those pixels in the ``template" PSF. This is done to maintain as much stellar PSF structure as possible in the reference image, since in the vast majority of datasets the stellar PSF is the main source of signal at all locations of the field of view, including at large separations. This replacement strategy was chosen, therefore, to minimize biases in KLIP from deformations in the stellar PSF.

Given that the PSF depends on the stellar spectrum and the deformations of the JWST primary mirror (among other sources of optical aberrations), we split all the survey targets into five sets: one set for GO 4050 observations, two for GO 5835 observations and two for SURVEY 6005 observations. For each of the programs GO 5835 and SURVEY 6005, observations were split into two sets based on whether they were observed before or after January 28th 2025. This is because the wavefront sensing measurement on that day shows a major micrometeorite impact. This impact noticeably deformed the JWST primary mirror in all subsequent observations relative to previous ones. Observations from GO 4050 form a single set because they were all observed within a few days of each other (so there is little wavefront error variation), and they all have similar spectral types. We take the pixel-wise median across contaminant-free observations in every set to obtain ``template" PSFs per set, which are used to mask faint contaminants in the rest of the observations within that set. 

Across all observations that were masked using these templates, the total number of masks across all observations in the F444W filter is 349, and 164 in the F200W filter. For each template made from contaminant-free observations, the number of integrations used per template varies from 14 to 91; see Appendix \ref{s:app_masking} for details.

\begin{figure*}
    \centering
    \includegraphics[width=1\linewidth]{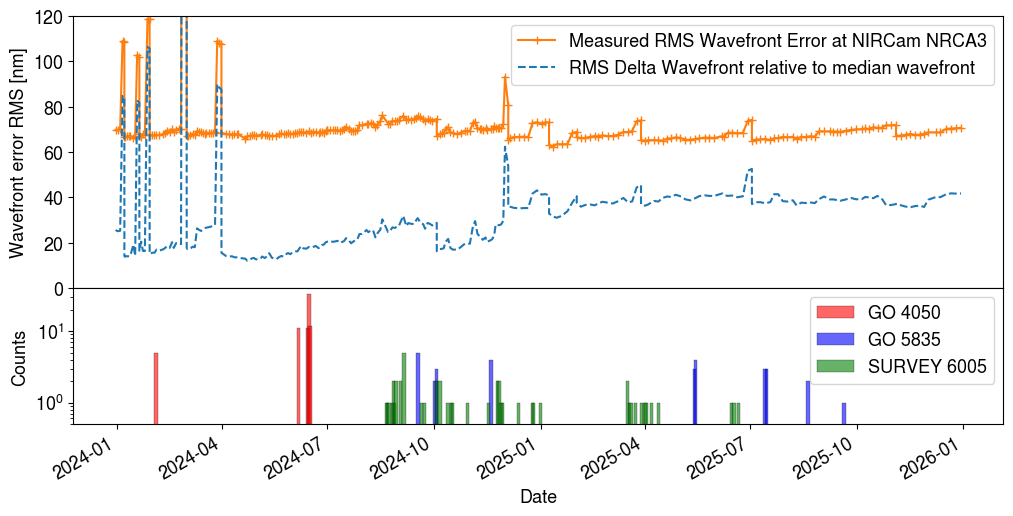}
    \caption{The wavefront error variation across the observations in our library contributes to the PSF diversity in our PSF library. Top: JWST NIRCam wavefront error measurements, and the deviations from the median value in this time interval. We plot the root mean square (RMS) of the wavefront error: this can be used as a metric to quantify the optical stability of the primary mirror. Lower RMS values indicate better alignment in the primary mirror (e.g., smaller-amplitude deformations or smaller tilts in mirror segments). The shape of the PSF in any given observation depends on the wavefront error during that observation \citep{perrin2018updated}.  Bottom: Observations in our library, binned by observation date. Each bin is a different day. Targets observed with $N$ dithers are counted $N$ times. Observations where both filters were included in the library are only counted one time. The $y$-axis is shown in logarithmic scale to make bins with one single observation more visible.}
    \label{fig:obs_wfe}
\end{figure*}

All observed targets included in the PSF library are listed in Table \ref{tab:properties} and their properties are visualized in Figure \ref{fig:histograms}. Figure \ref{fig:obs_wfe} shows the date ranges of the observations in our library, together with the wavefront error variation in JWST measured in flight.

\section{Tuning KLIP parameters for optimal sensitivity}
\label{s:contrast_opt}

With the PSF library assembled, we now examine the contrast sensitivity achieved under several uses of the PSF library. In particular, we assess what combination of hyper-parameters for the KLIP algorithm results in the best sensitivity with such a large collection of reference images. We only do these tests for observations in the F444W filter, since we expect similar results from the F200W filter.

KLIP has three main tunable hyper-parameters when performing PSF subtraction: 

\begin{enumerate}
    \item The number of frames in the PSF library to compute principal components from. This is done by computing the Pearson correlation of every frame in the library relative to the science image, and only using the $N$ frames with the highest correlation to the science target for PSF subtraction. Intuitively, this implicitly filters out less suitable references due to different stellar spectral types or realizations of wavefront error. Empirically, library down-selecting via the Pearson correlation has shown positive effects on PSF subtraction \citep{stasevic2025optimising}. In \texttt{pyKLIP}, the number of frames to keep corresponds to the \texttt{maxnumbasis} hyper-parameter.
    \item The number of principal components to use in the PSF model (referred to as the number of Karhunen–Loève bases, or KL modes). This encodes the ``complexity" of the PSF model built from the available frames in the library.
    \item The number of sub-regions (concentric annuli) to split the science image into over which the principal components are computed. This accounts for different combinations of KL modes producing better sensitivity in different regions of the science image.
\end{enumerate}


To determine the hyper-parameters that deliver optimal sensitivity, we perform a coarse search of parameter space and compute the sensitivity achieved in each case. After preliminary testing, we found that \texttt{maxnumbasis} and number of annuli have a relatively small impact on contrast compared to the choice of KL modes. For this reason, and to save computation time, we chose a finer grid search in KL modes compared to the other two parameters, for which we choose a small number of representative values. Our results in the next section further validate this choice. In summary:

\begin{itemize}
    \item We vary the number of frames to compute principal components over (the \texttt{maxnumbasis} argument in \texttt{pyKLIP}). We test using the 100 and 300 most correlated frames with the science target, and also using all the frames in the PSF library ($\sim500$). This is to see if restricting KLIP to the $N$ most correlated frames improves contrast by avoiding ``contamination" of the KL modes with less correlated frames.
    \item We vary the number of KL modes from 1 to the maximum (one can only use as many KL modes as frames are available; that is, number of KL modes $\leq$\texttt{maxnumbasis}), specifically testing 1, 5, 10, 20, 40, 60, 80, 100, 200, 300, 400 and all ($\sim500$) KL modes. This is to see if there is an optimal amount of complexity needed in our PSF model, considering that higher complexity might also result in more planet signal loss.
    \item We test splitting the images into 1 and 3 (equally spaced) concentric annuli, with ``1 annulus" corresponding to treating the entire image as a single region. This is to evaluate the impact of isolating regions with different noise properties, since at small separations (innermost annulus) residual starlight dominates, while at larger separations (outermost annulus) background or detector noise dominates.
\end{itemize}

For a robust grid search, we perform KLIP subtractions on every target of the PSF library itself as a science target (excluding that observation from the library), and we analyze the contrast achieved in each case. We decide to test our library on targets from the library itself because, by construction, these observations contain little to no contamination of other astrophysical sources that could bias the contrast performance estimation. 

For each realization of a KLIP subtraction, we compute the $5\sigma$ contrast sensitivity by taking the standard deviation of the residual at various annuli centered at the star location. We correct our sensitivity by the coronagraph mask throughput (at small separations, planets themselves are partially dimmed by the mask), by small sample statistics (due to limited noise samples at small angular separations; \citealt{mawet2014fundamental}), and by the KLIP algorithm throughput (the planet signal loss caused by the PSF subtraction process itself).

\begin{figure*}[t]
    \centering
    \includegraphics[width=1\linewidth]{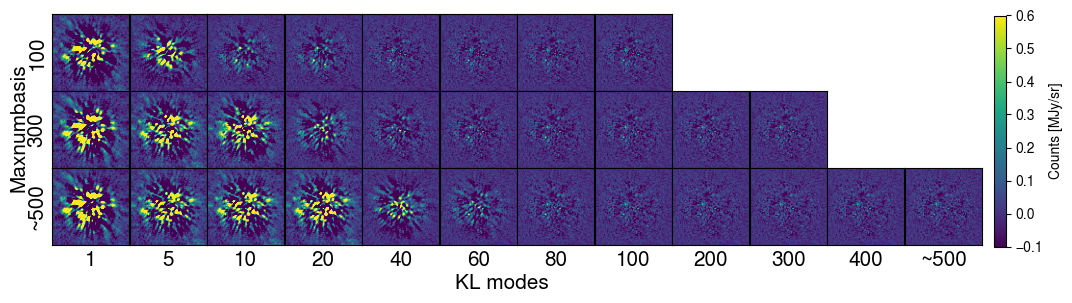}
    \caption{Illustration of PSF subtraction as a function of two KLIP hyper-parameters: number of KL modes and \texttt{maxnumbasis} (how many of the most correlated frames to compute principal components over). This is an arbitrary target in the F444W filter (2MASS J00135300-7441178) and, for brevity, only using the subtractions where the image is split into 3 annuli. The color scale is linear and identical for all plots. The exact value of \texttt{maxnumbasis=all} and \texttt{KL modes=all} varies slightly per target, but it is always $\sim500$. All panels show a field of view of $\sim 2.51\times2.51$ arcseconds and are oriented North up and East left. No synthetic planets are injected in these images. This is a visual illustration of the quantitative results shown across all targets in the following figures.\label{fig:psfsubgrid}}
\end{figure*}

\begin{figure}
    \centering
    \includegraphics[width=1\linewidth]{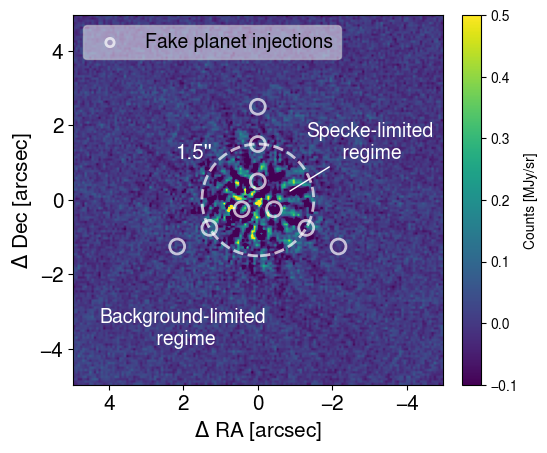}
    \caption{Schematic diagram of fake planet injections used for Figures \ref{fig:calcon05} and \ref{fig:calcon3} illustrated on an arbitrary target (AP Col, see Figure \ref{fig:ellsexamples}). We also show the two noise regimes discussed throughout the manuscript.}
    \label{fig:fake_schem}
\end{figure}

To account for algorithm throughput, we perform  injection-recovery tests of synthetic planet signals using planet PSF models from \texttt{STPSF}. Importantly, at separations beyond $\sim$2 arcseconds from the star most targets in the F444W filter are already in the ``background limited regime" (where no significant residual starlight is left, and where sensitivity is mostly influenced by the depth of the science observation), and within $\sim$1.5 arcseconds most targets are in the ``speckle limited regime" or ``contrast limited regime" (where sensitivity is limited by the quality of the PSF subtraction). To account for this, we inject synthetic planets at separations of 0.5, 1.5 and 2.5 arcseconds from the star when correcting for planet signal loss. At each separation we inject these planets at three different position angles: 0, 120 and 240 degrees East of North. This is done to account for azimuthal asymmetries in the PSF subtraction residuals. The median signal loss per separation is used to calibrate our sensitivity to real exoplanet detections. Figure \ref{fig:fake_schem} illustrates the two noise regimes we discuss in the next section and throughout the manuscript, along with the locations of the fake planet injections for our analysis in Section \ref{s:results_grids}. Our choice of throughput calibration with three separations and three position angles per separation is to avoid long computation times given the large number of targets in our PSF library. Our results in Section \ref{s:results_grids} are not strongly sensitive to the number of injected fake planets (see Appendix \ref{s:app_algothroughput}).

\section{Results: Optimizing contrast from KLIP hyper-parameters}
\label{s:results_grids}

Figure \ref{fig:psfsubgrid} shows the PSF subtraction quality varying as a function of two KLIP hyper-parameters: the number of principal components to include in the PSF model (KL modes) and number of frames to compute principal components over (\texttt{maxnumbasis}). It is clear that higher numbers of KL modes generally result in cleaner PSF subtractions. Similarly, the subtraction with 10 KL modes is better for \texttt{maxnumbasis=100} than with \texttt{maxnumbasis=all} ($\sim500$). This is expected, since by definition \texttt{maxnumbasis=100} includes only the 100 most correlated (or most similar) frames to the science image, so at equal numbers of KL modes, the average quality of references is higher. Nevertheless, having more KL modes available, the subtraction with \texttt{maxnumbasis=all} quickly catches up when using more principal components and produces comparable or better contrast than \texttt{maxnumbasis=100}.

However, Figure \ref{fig:psfsubgrid} only illustrates the ``raw" contrast achieved with PSF subtraction, and does not account for planet signal loss or ``over-fitting". While two subtractions may look comparable by eye, one may be able to keep more planet signal (that is, have a higher algorithm throughput) than the other, resulting in better detection sensitivity overall. The injection-recovery tests with synthetic planets described in Section \ref{s:contrast_opt} are meant to account for these signal loss differences. These tests result in calibrated contrast sensitivity curves. 

\begin{figure*}[t]
    \centering
    \includegraphics[width=1\linewidth]{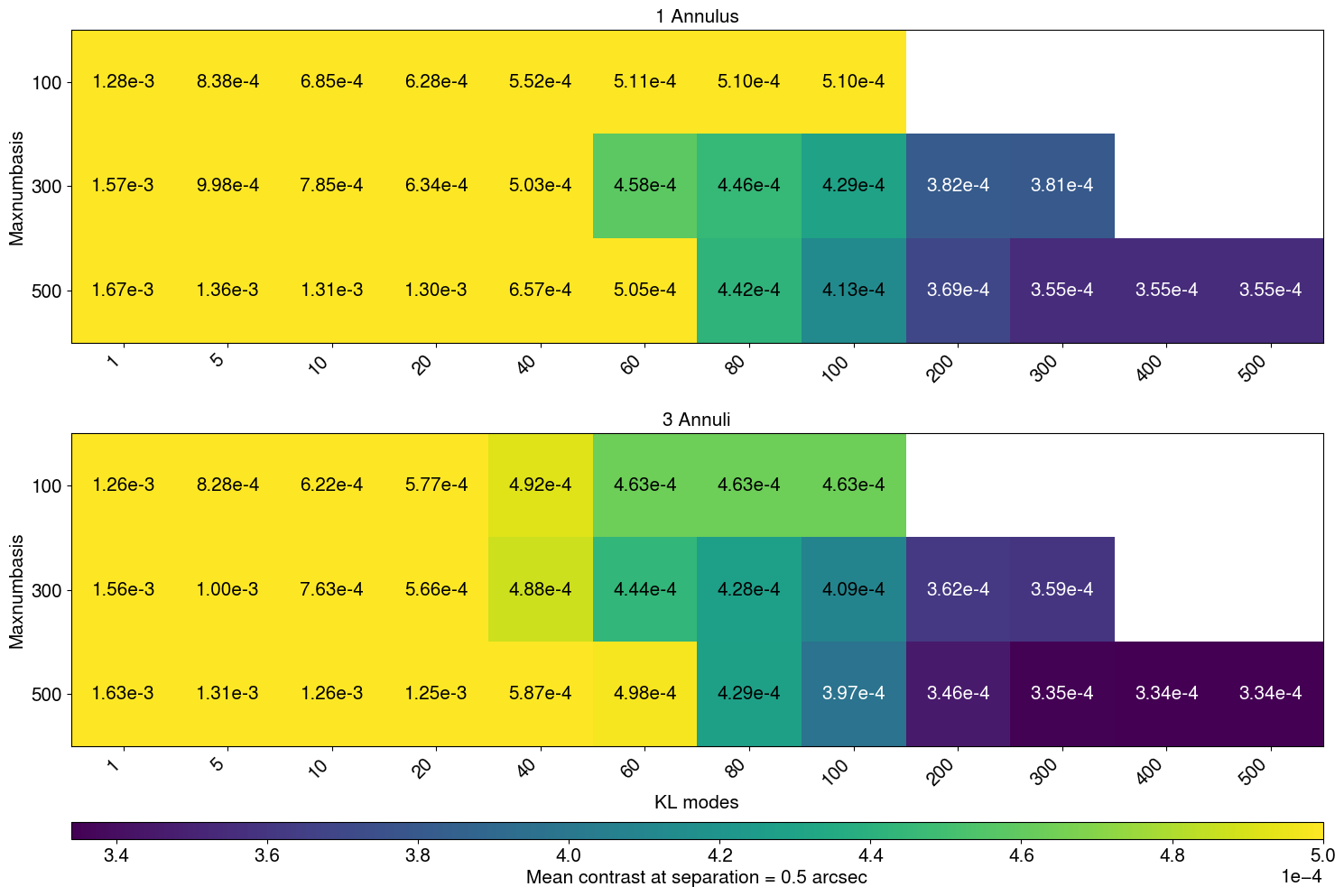}
    \caption{Mean 5$\sigma$ contrast limit at 0.5 arcsec across targets in our library, as a function of KLIP hyper-parameters. The top panel uses the full-frame images during KLIP (effectively one annulus), while the bottom splits each frame into 3 concentric annuli. At close separations, using higher KL modes and 3 annuli results in the best sensitivity. Each panel shows the mean 5$\sigma$ contrast sensitivity (see e.g., Figures \ref{fig:twa23} and \ref{fig:ellsexamples}) evaluated at separation of 0.5 arcseconds from the host star, calibrated with injection-recovery tests. Lower values indicate deeper and better sensitivity. There is a clear systematic improvement in sensitivity from using 3 annuli as opposed to only 1, and there is a preference for using many KL modes (almost as many as are available). However, the difference between using, for instance, 400 and 500 KL modes is rather small (less than $\sim2\%$ difference, while going from 1 annulus to 3 annuli at 500 KL modes results in a $\sim10\%$ gain in contrast). This is expected from the ``speckle" or ``contrast limited regime".}
    \label{fig:calcon05}
\end{figure*}

\begin{figure*}[t]
    \centering
    \includegraphics[width=1\linewidth]{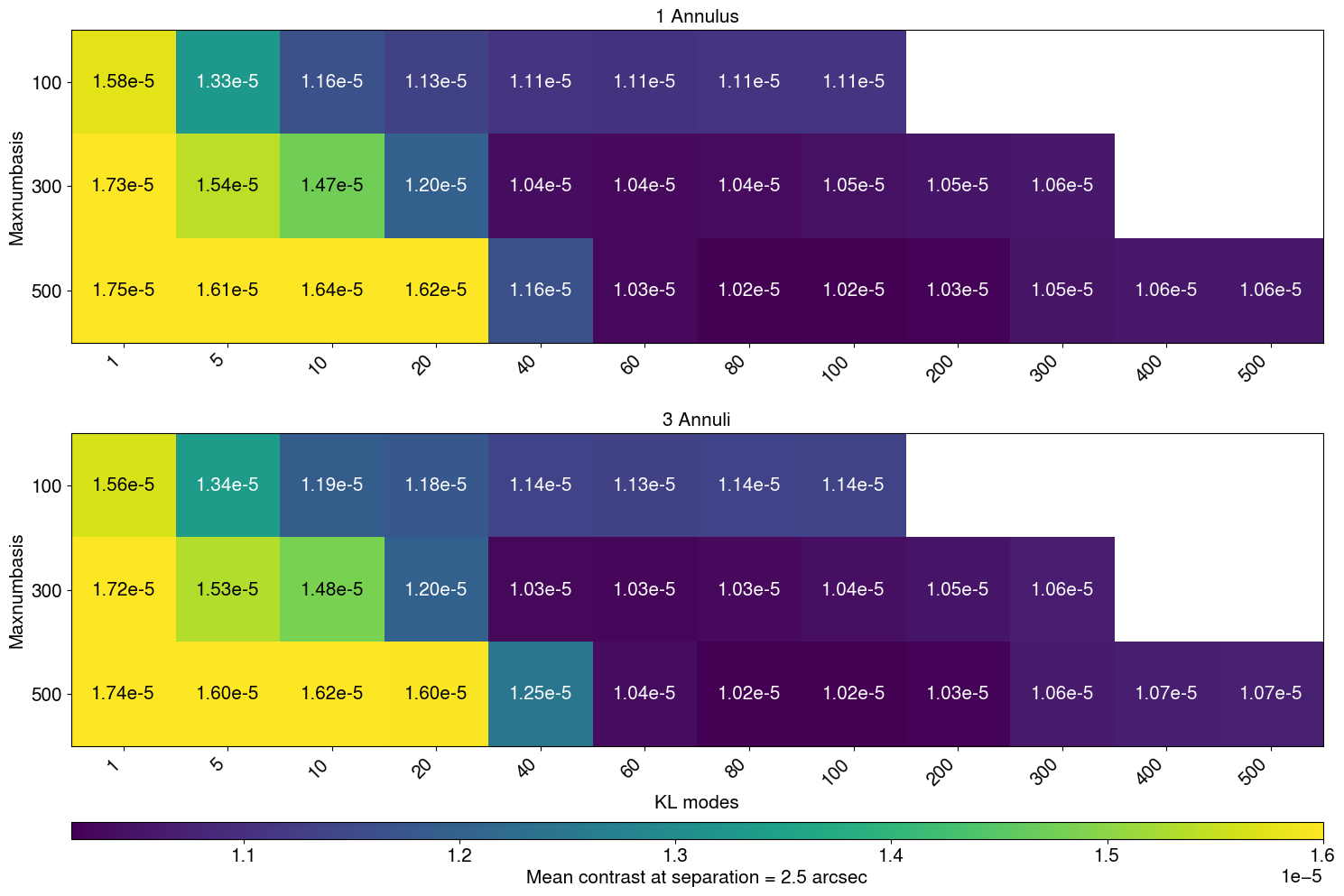}
    \caption{Same as Figure \ref{fig:calcon05}, but measuring contrast at 2.5 arcsec. Lower values indicate deeper and better sensitivity At larger separations, using a moderate number KL modes shows best sensitivity, with 1 and 3 annuli showing more similar performance. Unlike very close separations ($<1.5$ arcsec), at these slightly wider separations a smaller number of KL modes ($\sim 80 - 200$) tend to yield superior sensitivity, though the difference is not large (going from 50 to 300 KL modes changes contrast by less than $\sim3\%$). At large separations, the systematic difference between using 1 or 3 annuli is also reduced, even showing a small boost in contrast ($\sim2-3\%$) when using 1 annulus rather than 3. This is likely because most targets at this separation are in the ``background limited regime", where brighter reference stars with lower background noise likely improve sensitivity.}
    \label{fig:calcon3}
\end{figure*}

 With this in mind, we compare the calibrated contrast at very close separations ($0.5$ arcsec)\footnote{We find that our injection-recovery-calibrated contrast at $1.5$ arcsec shows qualitatively similar results to $0.5$ arcsec in terms of preferred KLIP hyper-parameters, so for brevity we only discuss explicitly the contrast at $0.5$ arcsec.} and at larger separations ($2.5$ arcsec) from the host star. We perform these tests on all targets in the PSF library, and we illustrate the results in Figures \ref{fig:calcon05} and \ref{fig:calcon3}. Throughout the rest of the manuscript, all contrast limits are calibrated with injection-recovery tests.

\subsection{Contrast in the speckle-limited regime}
\label{s:speckleregime}

At very close separations, we find that two main choices result in the deepest contrast (and therefore the best sensitivity): using 3 annuli rather than 1, and using a relatively high number of KL modes in the PSF model (above $\sim300$). Figure \ref{fig:calcon05} shows the median $5\sigma$ contrast sensitivity at $0.5$ arcseconds as a function of KLIP hyper-parameters. At $0.5$ arcseconds of angular separation, this is representative of the contrast in the ``speckle-limited regime", where the exoplanet detection sensitivity is mainly driven by the quality of the PSF model. Intuitively, using as wide a diversity of PSF references gives KLIP the greatest flexibility to model complicated features of the stellar PSF, resulting in superior sensitivity (see Section \ref{s:discussion}).

At these close separations, other parameter choices are less significant. The choice of the \texttt{maxnumbasis} parameter affects less strongly the contrast achieved when the number of KL modes is on the high end (the best sensitivity at \texttt{maxnumbasis=300} and with \texttt{maxnumbasis=all} only differ by $\sim2\%$). Nevertheless, the best sensitivity is typically found when using \texttt{maxnumbasis=all}.  Similarly, the exact choice of number of KL modes above a certain number does not strongly affect the overall contrast (at \texttt{maxnumbasis=all}, the sensitivity at 400 KL modes and above only changes by less than $\sim2\%$ overall).

\subsection{Contrast in the background-limited regime}
\label{s:backregime}

At larger separations, we find a qualitatively different behavior. This corresponds to the ``background-limited regime". Figure \ref{fig:calcon3} shows the contrast sensitivity evaluated at $2.5$ arcseconds. In this case the preference for 3 annuli over 1 annulus is no longer apparent. In fact, the deepest contrasts are found when using 1 annulus, but the difference with 3 annuli is small: at $\sim50$ KL modes and above, going from 1 to 3 annuli changes the sensitivity by typically by $2-3\%$ or less. 

Interestingly, we now see a marginal preference for a moderate number of KL modes, around $\sim80-200$.Two main factors could explain the preference for moderate numbers of KL modes. The first factor is that higher numbers of KL modes are likely to increase planet signal loss. While at close separations the gain in speckle suppression of high KL modes may outweigh planet signal loss, at larger separations this may no longer be true since there are no speckles to suppress. The second factor involve the level of background noise itself. Intuitively, the sensitivity in this regime depends on the observation time and the brightness of the science target. However, KLIP introduces a dependence on the background level of reference targets in the library. On one hand, references with much shorter integration times will have higher background noise. This can increase the background noise in the science target after PSF subtraction. On the other hand, part of the KLIP process is mean-scaling the references to match the brightness of the science target. As a result, the background noise can be scaled up during this process if the reference is much fainter than the science target (see also Section 3.1 in \citealt{BowensRubin_2025}). Considering the wide range of star brightnesses and integration times in the PSF library, it is likely that a combination of both factors contribute in different levels to the marginal reduction in sensitivity at large separations. Future work could test selecting within the library only the observations with the most similar stellar brightness or the highest background SNR to minimize the background noise injected during KLIP, if the goal is to truly maximize sensitivity specifically in the background-limited regime.

\section{Sensitivity improvements from our PSF library}
\label{s:bigcomparison}

In this section we investigate the sensitivity improvements obtained from using our large PSF library in PSF subtraction. We examine three scenarios, representing three possible use cases for our library:

\begin{itemize}
    \item \textit{PSF library and a dedicated reference star}: If an observation already has a dedicated dithered reference star, we assess whether using our PSF library in addition to the reference star improves sensitivity.
    \item \textit{PSF library and self-referenced surveys}: If an observation was taken as part of an uninterruptible sequence for self-referencing, we assess whether using our PSF library in addition to the contemporaneous observations improves sensitivity.
    \item \textit{PSF library with and without contemporaneous references}: In the context of planning future observations, we investigate whether it is still necessary to request contemporaneous reference stars (either dedicated stars or as self-referencing), or if the PSF library is sufficient for optimal sensitivity.
\end{itemize}

We present the results our analyses below. Observations processed with RDI (the ones with a single roll angle) consist simply of applying KLIP on them using the reference images of other stars to construct the KL basis. We also analyze archival observations with two science roll angles, which were processed with ADI+RDI. For ADI+RDI, both science rolls are applied KLIP independently computing the KL bases across the reference library of other stars and the other science roll, then both PSF-subtracted rolls are de-rotated and mean-combined.

In all cases, the raw, uncalibrated contrast curves are sampled using an annulus width of 2.8 pixels, corresponding approximately to one resolution element ($\lambda/D$) for the F444W filter with the 335R coronagraph in NIRCam. We calibrate algorithm throughput by injection and recovery of fake planets. Here we inject planets at 12 separations (from 0.25 arcsec to 3.0 arcsec in steps of 0.25 arcsec), and at 6 equally-spaced position angles per separation. We omit fake planets that land on top of other astrophysical sources (see Figures \ref{fig:twa23} and \ref{fig:ellsexamples}). The algorithm throughput at each separations is estimated by taking the median throughput of planet injections across position angles, and the throughput at intermediate separations is obtained with a linear interpolation. The calibrated contrast curves presented below are the raw contrast curves divided by the throughput.

\subsection{Using our library on targets with dedicated reference stars}
\label{s:fullvsdedicated}

Many exoplanet imaging observations are carried out with a dedicated reference star to enable PSF subtraction. The dedicated reference star is often a nearby bright star with similar spectral type as the science target observed right before or after the science sequence (to maximize similarity between reference and science PSF). Typically, the reference star is observed with a sub-pixel dither strategy \footnote{\url{https://jwst-docs.stsci.edu/jwst-near-infrared-camera/nircam-operations/nircam-dithers-and-mosaics/nircam-subpixel-dithers/nircam-small-grid-dithers}} to account for small mismatches in the position of the star behind the coronagraph mask \citep{soummer2014small,lajoie2016small}. We investigate the performance of our PSF library when added to the reference observation, and we compare it to the performance purely provided by the reference observation. 

We perform experiments on three targets. Two are taken from our PSF library itself: TWA 23 and TWA 46. We choose these because they have a dedicated dithered reference star also included in our library (HD 89063 served as reference for both targets). We process these observations using RDI. We additionally perform an experiment on an archival target: 14 Her \citep{bardalez2025jwst} (from JWST GO 3337, PI: Bardalez Gagliuffi, co-PI: Balmer). We choose 14 Her because it is one of the highest contrast observations taken with JWST/NIRCam to date and because in addition to a reference star it also includes an ADI roll angle (unlike TWA 23 and TWA 46). We process the 14 Her dataset following \citealt{bardalez2025jwst} \footnote{Detailed reproduction instructions can be found in \citealt{william_o_balmer_2025_15483953}}. We reproduce the published KLIP subtractions, with comparable detections of both the planet companion 14 Her c and the background galaxy interloper. For our KLIP subtractions, we use ADI+RDI, we use one annulus, \texttt{maxnumbasis=all} and all available KL modes ($\sim240$). We note that the authors use 25 KL modes in \cite{bardalez2025jwst}: we opt for all KL modes since they produce better contrast in our own reproduction. This difference in KL modes for 14 Her may be due to unidentified changes in the software versions of the used pipelines, since the starting data products are identical and the same data reduction steps are followed. Nevertheless, our reduction produces a comparable detection of both astrophysical sources in the field of view (the background galaxy and the companion 14 Her c) to the published image, showing comparable PSF subtraction performance.

\begin{figure*}[t]
    \centering
    \includegraphics[width=0.9\linewidth]{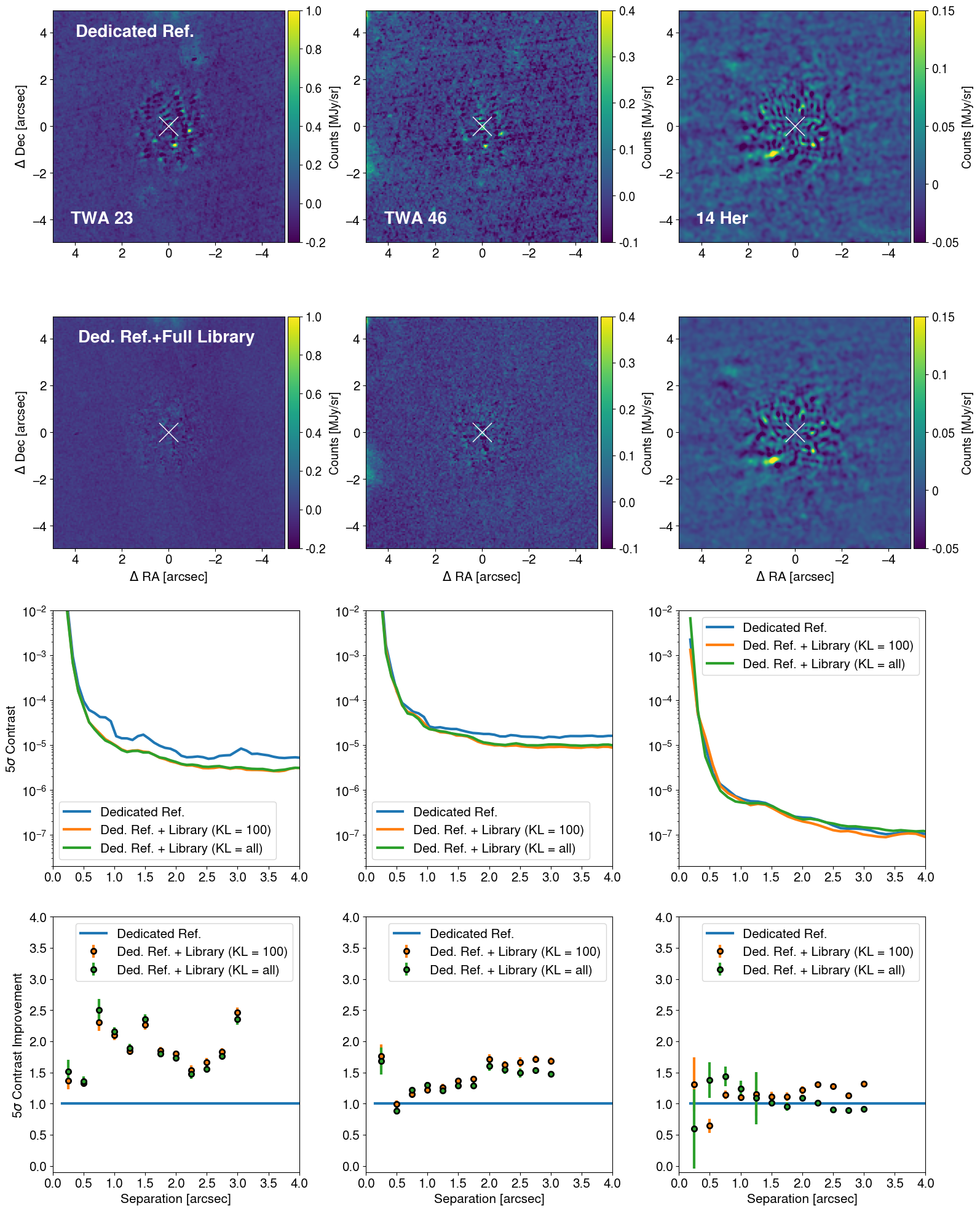}
    \caption{PSF subtraction using dedicated reference stars compared to using both reference stars and our large PSF library. Top: PSF subtraction with dedicated reference star (and an ADI roll for 14 Her). Second row: PSF subtraction incorporating our PSF library to the existing references. Third row: Contrast sensitivity curves. Bottom: Ratio between each contrast curve and the dedicated reference contrast curve, evaluated at the separations where fake planets were injected (error bars are propagated from the standard error on the median throughput, computed across fake planets per separation). The upper left panel shows artifacts from imperfect masking of off-axis sources (see Section \ref{s:selection}), visible as circular patches. All images are with all KL modes, except for the second 14 Her image, which uses 100 KL modes. A large PSF library along with a reference star can produce better sensitivity than the reference star alone.}
    \label{fig:twa23}
\end{figure*}

\begin{figure*}
    \centering
    \includegraphics[width=0.9\linewidth]{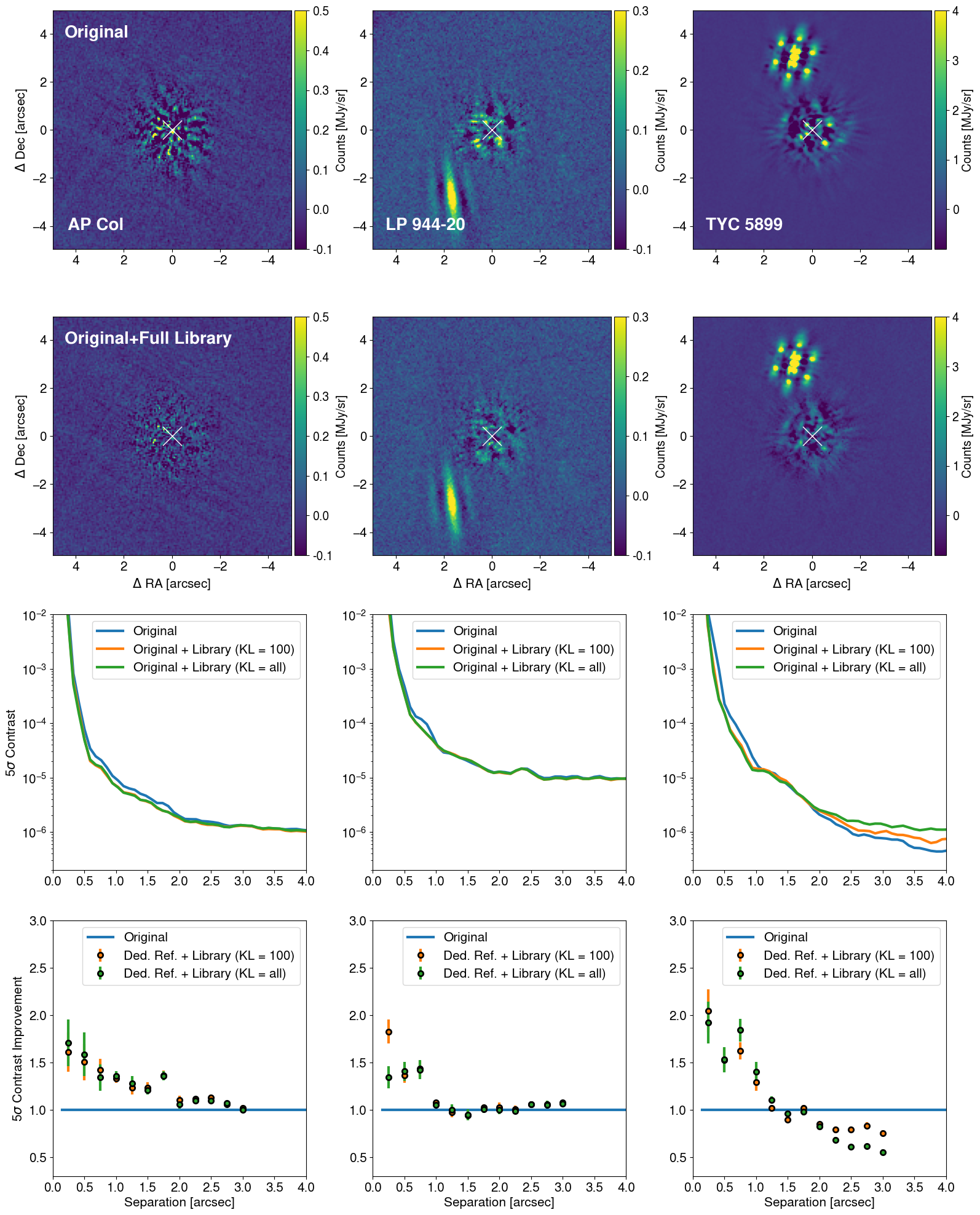}
    \caption{Same as Figure \ref{fig:twa23}, but on targets from program GTO 1184. Here, ``Original" refers to only using the small PSF library used originally in \cite{bogat2025probing}. ``Original + Full library" includes both the original PSF library in \cite{bogat2025probing} and our large PSF library. \texttt{KL modes = all} corresponds to $\sim500$. Adding a large PSF library improves sensitivity in the speckle-limited regime for targets of surveys not included in our library, but when used on bright stars or deep observations, it can diminish sensitivity in the background limit.}
    \label{fig:ellsexamples}
\end{figure*}

Figure \ref{fig:twa23} shows the results of these experiments. The contrast curves show that in all cases adding our library improves sensitivity, with the improvement being easily visible by eye in TWA 23 and TWA 46, and more marginal at close separations in 14 Her. In TWA 23 and TWA 46, interestingly, we note improvements both at close and large separations, with speckles being mitigated in the speckle-limited regime and noise being more Gaussian in the background regime. For TWA 23 and TWA 46, while the improvement in the speckle regime is driven substantially by speckle suppression from the PSF diversity in our library, it is also partly driven by suppression of background and detector noise. This is also the main driver behind the improvement in the background limit. The background and detector noise are mostly a consequence of the reference star observation of HD 89063. Since HD 89063 is much brighter stars than the science targets (see Table \ref{tab:properties}), it was observed for a shorter amount of time ($\sim126$ seconds per dither for HD 89063, while TWA 23 and TWA 46 were both observed for $\sim$1865 seconds each). Consequently, the short integration time of the reference star may contribute unwanted background and detector noise during PSF subtraction, such as the so-called 1/f noise\footnote{\url{https://jwst-docs.stsci.edu/known-issues/1-f-noise}} which is a likely contributor to the faint vertical (and some horizontal) stripes visible in the top row of Figure \ref{fig:twa23} for TWA 23 and TWA 46. These sources of noise are partially mitigated when additional references from our library with higher background SNRs are included, which drives the notable improvement in the background-limited regime in these two targets seen in Figure \ref{fig:twa23}, and likely contributes to the speckle-limited regime as well.

For 14 Her, we do not see a substantial difference in contrast when adding our large library. In some cases, we even see a decrease in sensitivity. This may be because in this particular dataset, the available reference stars already provide a very good starlight model that is hardly improved by the greater PSF diversity in our library. Additionally, using a larger number of KL modes might increase the risk of signal loss to a point where the greater starlight suppression does not actually result in better sensitivity (as was noted in Section \ref{s:backregime}; see also Appendix \ref{s:app_algothroughput}).

The bottom panels of Figure \ref{fig:twa23} show the relative difference in contrast. To assess the significance of the contrast improvements, at each separation where fake planets were injected we compute the standard error on the median throughput across fake planets. This error encapsulates the uncertainty on the median throughput (which is used as the calibration factor for our contrast limits), taking into account the throughput dispersion across injected planets due to, for instance, azimuthal differences in throughput. Our standard error on the median throughput is propagated to the contrast limits, and then to their ratios, which we show as error bars in the plot.

We see that for TWA 23 and TWA 46, adding our library improves sensitivity by up to $\sim2.5\times$ and $\sim1.7\times$, respectively. These differences are larger than the uncertainties due to throughput dispersion in most cases. On the other hand, we see that at close separations adding a large library to 14 Her does not significantly improve sensitivity in general, while it seems to show a small but significant improvement at large separations by up to $\sim1.3\times$.


\subsection{Using our library on archival self-referenced targets}
\label{s:outofdist}

Here we test our PSF library on data from an archival, published self-referenced survey. We choose three targets from the self-referenced program JWST GTO 1184 (PI: J. Schlieder) and analyzed in \cite{bogat2025probing}: AP Col, LP 944-20 and TYC 5899. All three observations were observed at two roll angles to enable ADI in addition to RDI. Both AP Col and LP 944-20 were observed as part of uninterruptible sequences, but TYC 5899 was observed months after the original sequences \footnote{TYC 5899 was originally observed as part of an uninterruptible sequence too, but the second roll angle observation failed. The observation we include here is a repeated two-roll observation.}. This provides two test cases of self-referenced targets with near contemporaneous (that is, near optimal) reference stars, and one test case where no close-in-time references are available. We assess the effect of our library in all of these cases.

We re-process both roll angles (ADI+RDI) of all three observations above in line with the details presented in the published reduction. In particular, we use the same KLIP hyper-parameters used in the paper: 6 KL modes and 4 annuli. Our reductions are presented in the top panels of Figure \ref{fig:ellsexamples}, and are consistent with the published reductions shown in Figure 13 of \cite{bogat2025probing}. For these subtractions, we use the same hyper-parameters as the published work because in our reduction, they produce comparable contrast limits as presented in the original work.

The second row panels of Figure \ref{fig:ellsexamples} shows the PSF subtraction achieved when incorporating our PSF library to the available reference stars from program GTO 1184. When including our large PSF library, just like in the rest of our experiments, we use 3 annuli and all KL modes, as this showed the best performance. Especially inside 1.5 arcsec, we see that our library was able to suppress residual starlight and remove several possible false positives, showing the benefits of using our library especially in the speckle-limited regime. In the color scale used, AP Col also shows a vertical stripe pattern (and a horizontal pattern to a lesser extent) due to detector noise, but it is also visible in our own reduction with a large PSF library. This suggests the pattern is intrinsic to the science target as opposed to a contribution of a single reference observation (as was the case in TWA 23 and TWA 46; see Section \ref{s:fullvsdedicated}), and that a large library cannot as easily remove these patterns.

This improvement is shown quantitatively in the panels of the bottom two rows of Figure \ref{fig:ellsexamples}. Inside $\sim1.5$ arcsec, using our PSF library brings improvements in sensitivity. We note how different aspects of these improvements are visible in the images compared to the contrast curves. The contrast curves encode the over-fitting-corrected sensitivity on average (something not encoded in the images), but the images show specific bright speckles being removed by our large library that could have been mistaken by planet signals (something not captured by the average sensitivity in the contrast curves). This serves as a reminder that the assumption of azimuthally-symmetric noise residuals in contrast curve calculations can be misleading regarding true sensitivity, and that image inspection and 2D sensitivity or signal-to-noise maps are valuable assessments. We highlight that given the correlated noise in starlight residuals, an important benefit of the PSF library can come from ruling out specific false positives (that is, removing persistent bright speckles), in addition to improving overall sensitivity. 

Another feature in Figure \ref{fig:ellsexamples} is that in the background limit, outside $\sim1.5$ arcseconds, using our PSF library does not significantly improve sensitivity, and it even worsens it in the deepest observation (TYC 5899). As discussed in Section \ref{s:backregime}, an explanation could be that the original PSF library in \cite{bogat2025probing} already has deep enough references to mitigate background noise (unlike in TWA 23 and TWA 46 in Figure \ref{fig:twa23}). Moreover, in TYC 5899 our library worsens contrast at large separations. As was also discussed in Section \ref{s:backregime}, a combination of higher planet signal loss and injected background noise could explain this difference. Particularly for the subtraction with 100 KL modes, injected background noise could be significant, given that TYC 5899 is significantly brighter than most stars in our library (W2=6.60 mag) and the combined observation is deeper than most observations in our library (the total exposure time for the two rolls was $\sim60$ minutes, while most of the PSF library is $\sim30$ minutes).

The bottom panels in Figure \ref{fig:ellsexamples} also show the ratio between contrast curves at the separations where fake planets were injected. The error bars are propagated from the standard error on the median throughput used to calibrate the raw contrast curves, in the same way it was done for Figure \ref{fig:twa23} (we also include our throughput correction plots in the Appendix; see Figure \ref{fig:allthroughputs}). We see that for all targets, we achieve statistically significant improvements in contrast at separations within $\sim1$ arcsec, corresponding to the suppression of specific starlight residuals. We also see that in the background limit, our large library achieves comparable performance to the original library. For TYC 5899, adding our library even decreases sensitivity significantly. This may be partly due to the bright off-axis star present in the field, which a large PSF library may have more flexibility to try to remove, but at the cost of a slightly worse suppression of background noise in general.

\subsection{Planning observations: Do we still need contemporaneous references?}
\label{s:contemporary}

Sections \ref{s:fullvsdedicated} and \ref{s:outofdist} show that given an existing observation, incorporating our large PSF library generally improves sensitivity at close separations. A natural question is the following: given our large PSF library, is it still necessary to have contemporaneous observations to supplement PSF subtraction? In other words, do single-target programs still require a dedicated reference star with dithers, and do multi-target programs still require scheduling uninterruptible sequences for self-referencing?

To explore this question, we compare the contrast sensitivity achieved in a few different ways:

\begin{enumerate}
    \item PSF subtraction via only ADI (when the science target has two contemporaneous rolls).
    \item PSF subtraction via only contemporaneous RDI (i.e., RDI using only contemporaneous observations to the science target, either from a dithered reference star or from self-referencing).
    \item PSF subtraction via contemporaneous RDI+ADI.
    \item PSF subtraction via only large library RDI (i.e., RDI using only our PSF library, with no contemporaneous observations).
    \item PSF subtraction via large library RDI and contemporaneous reference stars.
    \item PSF subtraction via large library RDI+ADI (i.e., RDI using our PSF library, and the two rolls of the science observation).
    \item PSF subtraction via large library RDI+contemporaneous RDI+ADI (i.e., RDI using our PSF library, the two rolls of the science observation and contemporaneous observations).
\end{enumerate}

We carry out these tests on AP Col, LP 944-20 (Figure \ref{fig:ellsexamples}) and 14 Her (Figure \ref{fig:twa23}) as illustrative examples. We choose these targets because they were observed with two roll angles and with contemporaneous observations in an uninterruptible sequences. 

\begin{figure*}
    \centering
    \includegraphics[width=\linewidth]{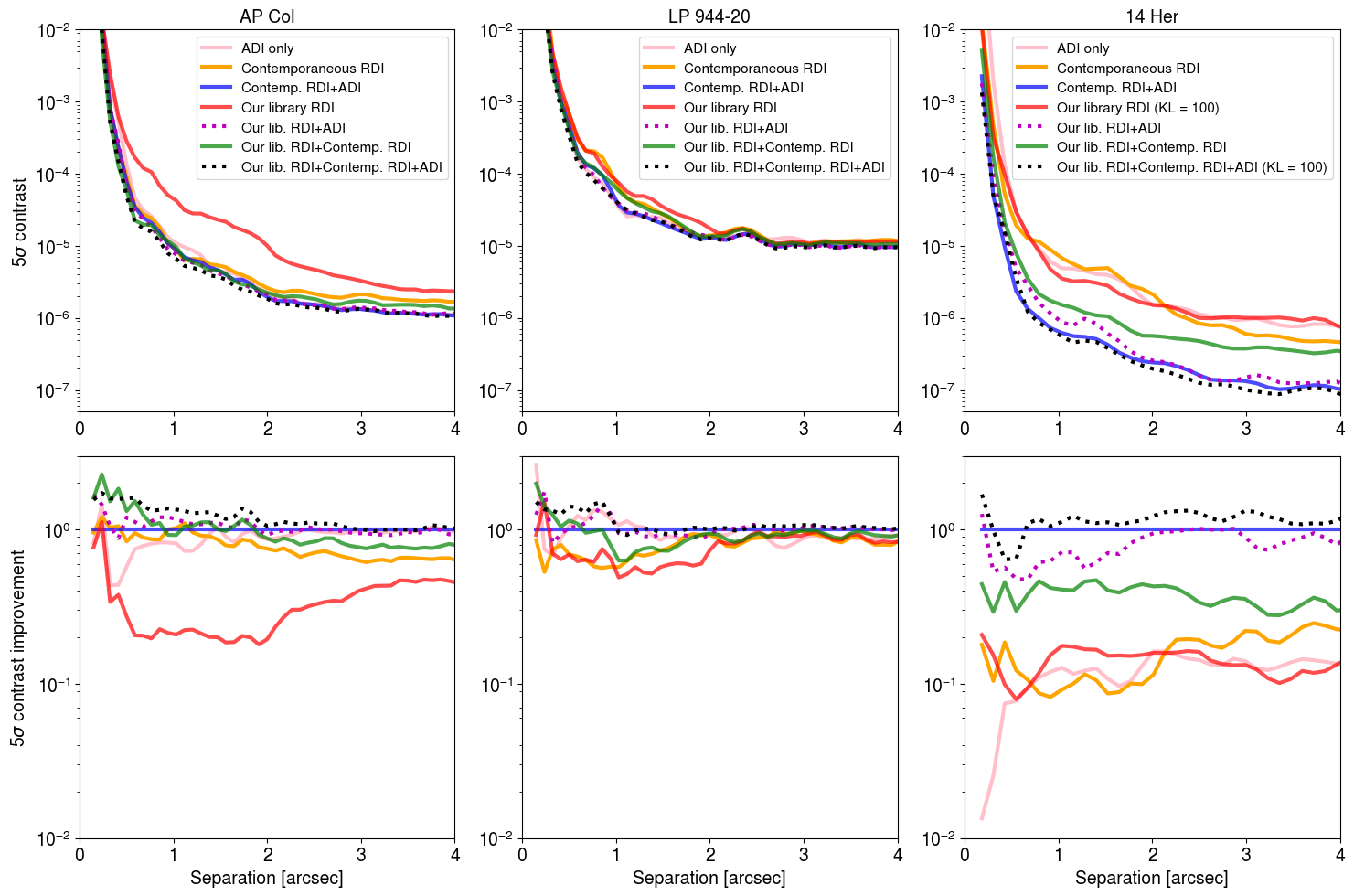}
    \caption{Comparison between contrast limits achieved with different sets of reference images. Top: 5$\sigma$ contrast limits for three targets. Bottom: Ratio of each curve to the nominal ``Contemporary RDI +ADI" curve. In the bottom panel, curves below the straight blue line indicate worse contrast sensitivity than the nominal contrast curve. Note that all panels are in logarithmic scale in the $y$-axis. Curves that do not include our library are obtained with the KLIP parameters noted in Sections \ref{s:fullvsdedicated} and \ref{s:outofdist}. All curves that include our library are using all KL modes, unless noted otherwise.}
    \label{fig:curves_combinations}
\end{figure*}

Figure \ref{fig:curves_combinations} show the contrast sensitivity curves for all seven experiments on all three targets. Using only our PSF library  yields close to the worst sensitivity in all cases. This clearly shows there is significant benefit to having contemporaneous reference observations, and that a large archival reference library is not enough to obtain similar sensitivity. In AP Col, using only our PSF library achieves $\sim2-5$ times worse sensitivity than contemporaneous references. In LP 944-20, using only our library with no contemporaneous references results in up to $\sim2$ times worse sensitivity. In 14 Her, using only our library results in $\sim5-10$ times worse sensitivity, but it achieves comparable sensitivity to only using ADI or only using contemporary RDI.

Figure \ref{fig:curves_combinations} further supports the conclusions from Figures \ref{fig:twa23} and \ref{fig:ellsexamples}: our library can result in statistically significant superior contrast (especially at close separations) when combined with contemporaneous reference stars and an ADI roll if available. 

It is worth highlighting that simply adding an ADI roll to our large library showed very significant improvements. This is particularly evident in the two brightest targets, AP Col and 14 Her. We find that our library and ADI together show up to $\sim2$ times better contrast relative to ADI only, and $\sim2-5$ times better relative to only using our library. For 14 Her, we find that our library and ADI together show $\sim6-10$ times better contrast relative to ADI only, or to only using our library.

This is in line with another feature in Figure \ref{fig:ellsexamples}: The first two columns (AP Col and LP 944-20) show visibly better starlight subtraction than the bottom row (TYC 5899), as noted by the amplitude of the residual speckles relative to the background. This is because AP Col and LP 944-20 were observed as part of uninterruptible sequences, while TYC 5899 was not. In other words, the ``Original" PSF library in \cite{bogat2025probing} includes observations contemporaneous to AP Col and LP 944-20, but not to TYC 5899. Importantly, this is the case even when including our own Full PSF library. This suggests that, even when a large library is available, contemporaneous reference observations (either from dedicated stars or in a self-referenced survey) significantly helps PSF subtraction.


\section{Discussion}
\label{s:discussion}

\textit{Limitations:} Our library shows contrast improvements in the speckle-limited regime (close angular separations) in most tested cases (Figures \ref{fig:twa23} through \ref{fig:curves_combinations}). However, as seen in Figures \ref{fig:ellsexamples} and \ref{fig:curves_combinations}, in the background regime our library can show negligible or even worse sensitivity. Without additional filtering (e.g., selecting only the brightest stars with the highest background SNR), this library is best used as a tool to improve contrast at small separations. Similarly, Figure \ref{fig:curves_combinations} shows that our library is not a replacement for contemporaneous reference observations, either from ADI rolls, dedicated stars with dithers, or self-referenced science targets. This is partly because, at the physical level, the performance of the library depends on the similarity between the PSFs in the library and the science target, which in turn crucially depends on factors such as the star centering behind the coronagraph, the stellar spectral type and the wavefront error in the telescope (Figure \ref{fig:obs_wfe}). The library is expected to perform better on average the more similarity there is between a science target and the images in the library in those aspects.

\textit{Too many KL modes?:} The preference for near-maximum number of KL modes in Section \ref{s:speckleregime} might be surprising given the apparent increased risk of planet over-fitting with a more complex PSF model. A few factors may be invoked to explain this. First, JWST data shows much lower variability and PSF diversity than that seen for ground-based data, due to the absence of atmospheric turbulence and the exceptional stability of the observatory \citep{telfer2024empirical}. As a result, even a complex PSF model with many KL modes has relatively limited flexibility. Secondly, ground-based data is most commonly taken using ADI, where the planet signal is present in all reference frames. A very high number of KL modes can more easily result in planet self-subtraction. The combination of lower PSF diversity and the absence of planet signals in RDI likely reduces the risk of planet over-fitting compared to ground-based data. Finally, targeted JWST observations often have dedicated reference stars: carefully chosen bright stars of similar spectral type as the science target, observed back-to-back with the science target and always with a sub-pixel dithering strategy. Our PSF library, on the other hand, is comprised of a large set of stars with diverse spectral types, observed across many months and predominantly without sub-pixel dithering. With such a stable observatory as JWST over $\sim$hours timescales, even a large set of reference stars may not be a cause of concern for over-fitting (compared to dedicated dithered reference stars), since there may not be enough speckle diversity to fit any arbitrary planet. While this is clearly observation dependent, this may also partially explain the preference for high numbers of KL modes (i.e., higher model complexity) in the speckle-limited regime, even when accounting for over-fitting.

\textit{Physically-motivated filtering of library:} Instead of filtering our library empirically based on the Pearson correlation (i.e., \texttt{maxnumbasis}), one can imagine filtering the library based on physical criteria. Some example criteria include star brightness, spectral type match, closeness in time, wavefront error realization and alignment behind the coronagraph. We do not experiment with them for two reasons. First, filtering based on a single physical criterion is likely not optimal. For example, a reference can be a good match in spectral type, but not a good match in wavefront error realization. A proper analysis on the effect of these criteria, then, may require a multi-dimensional parameter search that is outside the scope of this work. Second, the aggregate effect of these physical variables is reflected on the PSF structure as a whole. As such, the Pearson correlation criterion explored in this work should empirically capture most of these effects, albeit implicitly. Both of these reasons are supported by similar experiments with PSF libraries on ground-based observatories \citep{stasevic2025optimising}. Nevertheless, future work could investigate these effects in detail.

\textit{Practical recommendations:} Figures \ref{fig:calcon05} through \ref{fig:ellsexamples} suggest clear guidelines to maximize sensitivity in JWST NIRCam, given our PSF library. We summarize these recommendations as follows:

\begin{itemize}
    \item Even when contemporaneous reference observations are available, we  recommend adding our PSF library to increase sensitivity in the speckle-limited regime.
    \item Even with our full library being available, the best PSF subtraction in the speckle-limited regime is achieved when contemporaneous observations (e.g., from reference stars or uninterruptible sequences) are used in conjunction with our library. Therefore, we still recommend requesting contemporaneous references in future planning.
    \item When using our PSF library in the speckle-limited regime (here within $\sim1.5$ arcsec), we recommend using more than 1 annulus. We used 3 annuli in this work, but the advantage likely comes more from ``isolating" the speckle-limited regime in its own section of the image rather than from the specific number of annuli.
    \item In most cases, we recommend using close to all available KL modes when using our library to maximize sensitivity in the speckle-limited regime. 
\end{itemize}

\textit{Future steps:} Three main future directions can be followed from this work. First, we can employ our PSF library to enhance sensitivity in archival observations. Our findings suggest that adding our library improves contrast at close separations in the vast majority of cases. Re-visiting other JWST programs in the same observing modes with this expanded library has the potential to achieve deeper detection limits. Secondly, we can expand our library with archival observations, beyond those from programs GO 4050, GO 5835 and SURVEY 6005. We have seen that using a large number of KL modes is beneficial in the speckle-limited regime. It is likely that adding more data into our PSF library can improve sensitivity even more. Future work can investigate whether a larger library results in diminishing returns at any point, where either sub-selecting the library or using lower KL modes could be necessary. Finally, constructing PSF libraries for other observing modes in JWST could similarly increase high-contrast performance in those modes. These include other filters in JWST NIRCam, non-coronagraphic imaging with JWST NIRCam, or both coronagraphic and non-coronagraphic imaging with JWST MIRI. For MIRI, future work could investigate best practices for large PSF libraries in light of various detector effects and field-dependent wavefront aberrations.

\section{Conclusions}
\label{s:conclusions}

In this work we presented a PSF library to enhance sensitivity in high-contrast imaging observations, as part of a series of papers by the \textit{JWST Sub-Jupiters Survey Collaboration}. This library compiles data from JWST programs GO 4050, GO 5835 and SURVEY 6005, all observed with the NIRCam coronagraph in the F444W and F200W filters. We analyze the performance of the library in aiding PSF subtraction, in particular regarding the multiple hyper-parameters that can be tuned in the KLIP algorithm. We also directly compared the performance of the PSF library with both dedicated reference stars observed with a dithering strategy and with observations from other JWST surveys to assess the sensitivity improvements provided by this diverse PSF library. 

We summarize our findings, including guidelines for usage of the library, as follows:

\begin{itemize}
    \item Adding a large PSF library was found to enhance sensitivity in the speckle-limited regime in most cases, with small to moderate improvements ranging from $\sim$1.3$\times$ to $\sim2\times$. In other cases, the library has no significant effect on contrast (e.g., 14 Her in Figure \ref{fig:twa23} or at wide separations in Figure \ref{fig:ellsexamples}).
    \item While adding our PSF library can improve sensitivity at close separations, the best sensitivity is achieved when observations taken close in time to the science data were also available (Section \ref{s:contemporary}). Excluding contemporary references, our library alone consistently shows $\sim2-10\times$ worse contrast compared to including any contemporaneous reference observation.
    \item At close separations (in the speckle-limited regime), sensitivity is best improved with our library by using more than 1 annulus across the entire image (3 used in this work) and a high number of KL modes ($>300$) in the PSF subtraction in most cases (Figure \ref{fig:calcon05}). However, when planet signal loss is a bigger concern than speckle suppression, a lower number of KL modes may be optimal (Figure \ref{fig:twa23}).
    \item At larger separations (in the background-limited regime), the number of annuli was found to be less significant. Sensitivity was found to be best at a moderate number of KL modes ($\sim80-200$), though with relatively small differences compared to higher KL modes (Figure \ref{fig:calcon3}).
    \item In all cases, no systematic advantage was found by sub-selecting the most correlated frames in the PSF library before computing principal components (Figures \ref{fig:calcon05} and \ref{fig:calcon3}). For \texttt{pyKLIP}, this is reflected as the \texttt{maxnumbasis} parameter having relatively small effects. Albeit with small differences, the best contrast was consistently found to be with \texttt{maxnumbasis=all}, so we recommend this for all PSF subtractions.
\end{itemize}

We make this PSF library available to the high-contrast imaging community via Zenodo \footnote{\dataset[10.5281/zenodo.24119856]{https://doi.org/10.5281/zenodo.24119856}}.

\begin{acknowledgments}

This work is based on observations made with the NASA/ESA/CSA James Webb Space Telescope. The data were obtained from the Mikulski Archive for Space Telescopes at the Space Telescope Science Institute, which is operated by the Association of Universities for Research in Astronomy, Inc., under NASA contract NAS 5-03127 for JWST. These observations are associated with programs 4050 (PI: A. Carter), 5835 (PI: A. Carter), 6005 (PI: B. Biller), 1184 (PI: J. Schlieder) and 3337 (PI: D. C. Bardalez Gagliuffi, co-PI: W. Balmer). Support for programs 4050, 5835 and 6005 was provided by NASA through a grant from the Space Telescope Science Institute, which is operated by the Association of Universities for Research in Astronomy, Inc., under NASA contract NAS 5-03127. B.A.B. and B.J.S. acknowledge funding by the UK Science and Technology Facilities Council (STFC) grant nos. ST/V000594/1 and UKRI1196. A.S. acknowledges support from the National Science Foundation Graduate Research Fellowship under Grant No.~2139433. The JWST data presented in this article were obtained from the Mikulski Archive for Space Telescopes (MAST) at the Space Telescope Science Institute. The specific observations analyzed can be accessed via \dataset[doi:10.17909/v2m0-mn97]{https://doi.org/10.17909/v2m0-mn97}. This work benefited from the 2024 and 2025 Exoplanet Summer Program in the Other Worlds Laboratory (OWL) at the University of California, Santa Cruz, a program funded by the Heising-Simons Foundation. 
\end{acknowledgments}






\facilities{JWST}

\software{astropy \citep{astropy1, astropy2, astropy3}, matplotlib \citep{Hunter2007}, numpy \citep{harris2020array}, scipy \citep{2020SciPyNMeth}
          }

\begin{longrotatetable}
\begin{deluxetable*}{ccccccccc}
\tabletypesize{\small}
\tablewidth{\textwidth}
\tablecaption{Stars included in the PSF library \label{tab:properties}}
\tablehead{
\colhead{Target name} & \colhead{Spectral type}& \colhead{Sp. type ref.}&\colhead{K (mag)} & \colhead{K uncert. (mag)} & 
\colhead{W2 (mag)} & \colhead{W2 uncert. (mag)} & 
\colhead{F200W} & \colhead{F444W}
}
\startdata
$\ast$ 10 Crt & A2 & (1) & 5.235 & 0.023 & 5.044 & 0.061 & Y & Y \\
$\ast$ i Vel & B6 & (2) & 4.071 & 0.208 & 3.666 & 0.249 & Y & N \\
HD 101581 & K4.5V:k & (3) & 5.101 & 0.016 & ... & ... & Y & Y \\
CD-23 9765 & M2V & (4) & 6.097 & 0.023 & ... & ... & Y & Y \\
TWA 23 & M2.5Ve & (5) & 7.751 & 0.031 & 7.491 & 0.02 & Y & Y \\
TWA 46 & M3 & (6) & 9.232 & 0.023 & 8.892 & 0.02 & Y & Y \\
HD 89063 & M3D & (7) & 4.443 & 0.016 & 4.153 & 0.17 & Y & Y \\
TWA 33 & M4.75 & (1) & 9.118 & 0.023 & 8.459 & 0.02 & Y & Y \\
TWA 45 & M4.5 & (1) & 9.065 & 0.023 & 8.72 & 0.02 & Y & Y \\
TWA 35 & M4 & (8) & 8.723 & 0.024 & 8.334 & 0.019 & Y & Y \\
UCAC3 133-139341 & M6$\gamma$e & (9) & 9.43 & 0.023 & 8.968 & 0.021 & Y & Y \\
TWA 30 & M4.5Ve & (5) & 8.765 & 0.021 & 8.442 & 0.02 & Y & Y \\
TWA 36 & M4.8 & (1) & 9.847 & 0.019 & 9.453 & 0.02 & Y & Y \\
TWA 44 & M5$\gamma$e & (9) & 10.716 & 0.021 & 10.301 & 0.021 & Y & Y \\
V* Y Crt & M5e & (10) & 4.818 & 0.018 & 3.939 & 0.176 & Y & Y \\
TWA 26 & M9$\gamma$ & (9) & 11.503 & 0.023 & 10.819 & 0.02 & Y & N \\
TWA 29 & M9.5$\gamma$ & (9) & 13.369 & 0.036 & 12.652 & 0.022 & Y & N \\
IRAS 13041-5653 & M9D & (7) & 5.956 & 0.026 & 5.558 & 0.025 & Y & Y \\
V* V1217 Cen & M1Ve & (5) & 8.053 & 0.029 & 7.953 & 0.02 & N & Y \\
SCR J1237-4021 & M2.5Ve & (5) & 8.515 & 0.02 & 8.274 & 0.02 & Y & N \\
HD 207043 & G5V & (3) & 6.05 & 0.021 & 5.954 & 0.042 & Y & Y \\
V* V1005 Ori & M0 & (2) & 6.261 & 0.017 & 6.059 & 0.032 & Y & N \\
BD-13 6424 & M0 & (2) & 6.569 & 0.018 & 6.372 & 0.024 & Y & Y \\
2MASS J19300396-2939322 & M4 & (2) & 9.253 & 0.021 & 8.945 & 0.02 & Y & Y \\
SCR J0017-6645 & M3.0Ve & (5) & 7.703 & 0.021 & 7.498 & 0.02 & Y & Y \\
GJ 2006 A & M3.5 & (2) & 8.01 & 0.03 & 7.67 & 0.019 & Y & N \\
RX J0613.2-2742 & M3.8 & (2) & 7.145 & 0.024 & 6.822 & 0.02 & Y & Y \\
PM J00502+0837 & M5 & (2) & 8.897 & 0.021 & 8.539 & 0.02 & Y & Y \\
SCR J2033-2556 & M4.5Ve & (11) & 8.877 & 0.019 & 8.444 & 0.021 & Y & Y \\
PM J21214-6655 & M0 & (12) & 7.009 & 0.023 & 6.926 & 0.021 & Y & N \\
BPS CS 22898-0065 & M2 & (2) & 7.196 & 0.017 & 7.0 & 0.019 & Y & Y \\
RX J0019.7+1951 & M4 & (13) & 9.864 & 0.018 & 9.57 & 0.021 & Y & Y \\
PM J19560-3207 & M0 & (2) & 7.846 & 0.021 & 7.735 & 0.024 & Y & N \\
Barta 161 12 & M4 & (2) & 8.078 & 0.033 & 7.798 & 0.02 & Y & N \\
RX J0506.2+0439 & M4 & (2) & 8.067 & 0.024 & 7.74 & 0.02 & Y & Y \\
2MASS J19243494-3442392 & M4.6 & (2) & 8.785 & 0.025 & 8.41 & 0.02 & Y & Y \\
PM J22088+1144 & M4.3 & (13) & 9.037 & 0.021 & 8.673 & 0.021 & Y & Y \\
BD+05 378 & M0 & (14) & 7.069 & 0.031 & 6.936 & 0.02 & Y & Y \\
RX J0529.7-3239 & M4.5 & (2) & 8.316 & 0.027 & 7.928 & 0.021 & Y & Y \\
SIPS J1809-7613 & M5 & (2) & 8.989 & 0.021 & 8.544 & 0.02 & Y & Y \\
Smethells 20 & M1.5 & (2) & 7.854 & 0.024 & 7.709 & 0.021 & N & Y \\
GR* 9 & M5 & (2) & 9.278 & 0.021 & 8.883 & 0.021 & Y & Y \\
UCAC4 354-189365 & M5 & (2) & 7.552 & 0.017 & 7.161 & 0.019 & N & Y \\
V* AF Psc & M5 & (13) & 8.667 & 0.019 & 8.283 & 0.019 & Y & N \\
2MASS J09462782-4457408 & M5.5 & (12) & 9.992 & 0.021 & 9.56 & 0.02 & Y & Y \\
SCR J0336-2619 & M5.8 & (2) & 9.763 & 0.021 & 9.325 & 0.02 & N & Y \\
SIPS J2335-3401 & M6 & (2) & 10.758 & 0.021 & 10.266 & 0.02 & Y & Y \\
DENIS J004135.3-562112 & M7.8 & (2) & 10.864 & 0.019 & 10.366 & 0.02 & N & Y \\
V* WW PsA & M4 & (2) & 6.932 & 0.029 & ... & ... & Y & Y \\
MCC 124 & M1V & (15) & 6.261 & 0.023 & ... & ... & Y & Y \\
2MASS J00251465-6130483 & K7 & (2) & 7.749 & 0.026 & ... & ... & Y & Y \\
2MASS J00422033-7747397 & K3.5 & (2) & 7.529 & 0.031 & 7.494 & 0.019 & Y & Y \\
2MASS J02045317-5346162 & M3.8 & (2) & 9.563 & 0.021 & 9.223 & 0.02 & Y & Y \\
2MASS J03190864-3507002 & K7 & (2) & 7.723 & 0.023 & 7.607 & 0.02 & Y & Y \\
2MASS J03304909-4555573 & K3 & (2) & 7.103 & 0.023 & 7.102 & 0.02 & Y & Y \\
2MASS J08413264-6825403 & K7 & (12) & 7.379 & 0.018 & 7.324 & 0.02 & Y & Y \\
2MASS J21130526-1729126 & K6V & (16) & 7.583 & 0.018 & 7.497 & 0.022 & Y & Y \\
2MASS J21443012-6058389 & K8 & (2) & 7.874 & 0.026 & ... & ... & Y & Y \\
2MASS J12075091+1856558 & K2V & (17) & 7.028 & 0.023 & 7.058 & 0.02 & Y & Y \\
2MASS J05332558-5117131 & K7 & (2) & 8.158 & 0.023 & 8.061 & 0.02 & N & Y \\
2MASS J06434625-7158356 & G6 & (2) & 7.278 & 0.042 & 7.317 & 0.02 & Y & Y \\
2MASS J23274943-8613187 & G8 & (2) & 7.5 & 0.031 & 7.425 & 0.02 & Y & N \\
2MASS J05023042-3959129 & K4V & (18) & 8.05 & 0.023 & 8.076 & 0.02 & Y & Y \\
2MASS J04021648-1521297 & K3.4 & (2) & 7.573 & 0.02 & 7.564 & 0.021 & Y & N \\
2MASS J00135300-7441178 & G7 & (2) & 6.962 & 0.023 & 6.934 & 0.02 & Y & Y \\
2MASS J00345120-6154583 & K5 & (2) & 6.533 & 0.018 & 6.378 & 0.023 & Y & Y \\
2MASS J01232126-5728507 & G6 & (2) & 6.847 & 0.029 & 6.784 & 0.022 & Y & N \\
2MASS J01574896-2154052 & G4 & (2) & 6.472 & 0.026 & ... & ... & Y & N \\
2MASS J05203803-3945179 & F6 & (2) & 6.144 & 0.024 & 6.051 & 0.035 & Y & Y \\
2MASS J07210672+6739425 & K0V & (19) & 6.246 & 0.017 & 6.138 & 0.033 & Y & N \\
2MASS J07234358+2024588 & K3(V) & (20) & 6.879 & 0.018 & ... & ... & Y & Y \\
2MASS J07423606-5917507 & G2.5 & (21) & 5.701 & 0.021 & 5.606 & 0.059 & Y & N \\
2MASS J08555568+3611464 & G9 & (20) & 5.83 & 0.02 & 5.747 & 0.051 & Y & N \\
2MASS J12122084+5429086 & K7V & (22) & 6.059 & 0.017 & ... & ... & Y & N \\
2MASS J15353016+3612347 & K0V & (19) & 6.496 & 0.029 & 6.449 & 0.025 & Y & Y \\
2MASS J18130721+4128315 & F8(V) & (20) & 5.918 & 0.018 & 5.822 & 0.049 & N & Y \\
2MASS J21310171+2320074 & K8V & (15) & 6.382 & 0.026 & 6.238 & 0.026 & Y & Y \\
2MASS J21520973-6203085 & F6 & (2) & 6.027 & 0.021 & 5.937 & 0.042 & Y & N \\
2MASS J23001929-2609135 & G3V & (20) & 5.943 & 0.027 & ... & ... & Y & N \\
2MASS J23301372+3142210 & K2.5V & (23) & 6.145 & 0.021 & ... & ... & N & Y \\
2MASS J23561067-3903086 & K2V & (18) & 5.907 & 0.023 & ... & ... & N & Y \\
2MASS J03481148-7441388 & F3 & (2) & 6.136 & 0.018 & 6.03 & 0.035 & N & Y \\
2MASS J04213348-2543091 & F9V & (24) & 6.873 & 0.026 & 6.83 & 0.02 & Y & Y \\
2MASS J04384393-2702018 & G3 & (2) & 6.866 & 0.016 & 6.886 & 0.02 & Y & Y \\
2MASS J04480518-8046452 & F7 & (2) & 6.868 & 0.029 & 6.818 & 0.021 & Y & N \\
2MASS J05373962-2837346 & F4 & (2) & 6.277 & 0.02 & 6.186 & 0.027 & N & Y \\
2MASS J07370278-5221570 & F5V & (25) & 6.69 & 0.02 & 6.625 & 0.023 & Y & N \\
2MASS J06190805-0326203 & G0V & (20) & 6.552 & 0.02 & 6.448 & 0.024 & Y & N \\
2MASS J00065008-2306271 & F3 & (2) & 5.24 & 0.024 & 5.019 & 0.074 & N & Y \\
2MASS J01162926+4256219 & G0V & (19) & 5.177 & 0.016 & ... & ... & Y & Y \\
2MASS J01460105-2720556 & F0IV & (26) & 5.456 & 0.021 & ... & ... & N & Y \\
2MASS J04460056+7636399 & F7V & (20) & 5.284 & 0.02 & 5.116 & 0.072 & N & Y \\
2MASS J06223097-6013072 & G1V & (18) & 5.045 & 0.018 & 4.878 & 0.094 & Y & Y \\
2MASS J07330059+3701475 & G5V & (19) & 5.589 & 0.023 & 5.46 & 0.059 & Y & Y \\
2MASS J23030499+2055068 & G1V & (19) & 5.148 & 0.02 & 4.938 & 0.085 & Y & Y \\
2MASS J04152578+0611588 & G6V & (23) & 5.271 & 0.018 & 5.095 & 0.028 & Y & N \\
2MASS J04152879+0611128 & G0IV & (27) & 4.903 & 0.02 & 4.766 & 0.085 & Y & N \\
2MASS J06064047+1532317 & K0V & (19) & 4.822 & 0.017 & 4.748 & 0.097 & Y & Y \\
2MASS J19341977+5114121 & F7V & (19) & 4.494 & 0.018 & 4.203 & 0.143 & N & Y \\
2MASS J20395460+0620118 & K7Ve & (28) & 7.133 & 0.017 & 7.05 & 0.02 & Y & N \\
\enddata
\tablecomments{The reference for all K band magnitudes is \cite{Vizier_2mass}. All W2 magnitudes are from \cite{cutri2021vizier}. Some observations are only included in one of the filters (F444W or F200W) due to contamination being present in one filter or the other only. References for spectral types are: (1) \cite{2023AJ....165..269L}, (2) \cite{2024AJ....168..159L}, (3) \cite{2006AJ....132..161G}, (4) \cite{1986AJ.....92..139S}, (5) \cite{2017ApJ...840...87R}, (6) \cite{2006AJ....132..866R}, (7) \cite{macconnell2010vizier}, (8) \cite{2015ApJS..219...33G}, (9) \cite{2017ApJS..228...18G}, (10) \cite{1954ApJS....1..175B}, (11) \cite{2018AJ....156...49R}, (12) \cite{2019AJ....157..234S}, (13) \cite{2017AJ....154...69S}, (14) \cite{2007ApJ...662.1254S}, (15) \cite{2013ApJ...762...88M}, (16) \cite{2014ApJ...788...81M}, (17) \cite{simbad}, (18) \cite{Torres_2006}, (19) \cite{Tabernero_2017}, (20) \cite{2007AJ....133.2524W}, (21) \cite{2006ApJ...649L.115Z}, (22) \cite{LopezSantiago_2010}, (23) \cite{2003AJ....126.2048G}, (24) \cite{1964PLPla..28....1J}, (25) \cite{1975mcts.book.....H}, (26) \cite{2014MNRAS.443.2815P}, (27) \cite{Ammler_2009}, (28) \cite{2010AJ....140..119S}. }
\end{deluxetable*}
\end{longrotatetable}

\appendix

\section{Data reduction details}
\label{s:app_datared}

The data reduction consists of four major steps: JWST pipeline stage 1, JWST pipeline stage 2, PSF subtraction preparation, and PSF subtraction. 

We apply the standard JWST stage 1 pipeline, except we skip the interpixel capacitance correction (\texttt{ipc}) and dark subtraction (\texttt{dark\_current}) steps. For ramp fitting, we originally used the \texttt{LIKELY} algorithm \cite{brandt2024a,brandt2024b}, but we found that the sharpness of the NIRCam PSF caused incorrect flagging of the PSF core in many cases, so we opted for the default \texttt{OLS\_C} algorithm.  We additionally apply a 1/f noise correction with a median filter. We apply the stage 2 of the JWST pipeline without any custom modifications. 

PSF subtraction preparation consists of median frame subtraction, bad pixel cleaning, PSF re-centering, background subtraction, masking of neutral density squares and frame padding. 

For bad pixel cleaning, we apply an iterative sequence of pixel flags and corrections:

\begin{enumerate}
    \item First, we flag all pixels identified through the data quality (DQ) array from the stage 1 pipeline. We clean these pixels with a 2D interpolation with kernel size of 7.
    \item Next, we find pixels that are more than 6$\sigma$ variable across different integrations in a given dataset. We clean these pixels by substituting them by the median value across integrations in that pixel.
    \item Next, we flag pixels using sigma clipping with 5$\sigma$ within a 5$\times$5 pixel box centered at each pixel. We clean these pixels with a 2D interpolation with kernel size of 7.
    \item Next, we perform a more customised round of sigma clipping with a 6$\sigma$ threshold within a 3$\times$3 pixel box, excluding diagonally adjacent pixels from the outlier calculation. To avoid overflagging of true astrophysical sources, we exclude pixels close to the PSF core (\texttt{mask\_psf=True}), adjacent clusters of at least 6 identified pixels (\texttt{mask\_cluster\_size=6}), and pixel clusters within 6 pixels of each other (\texttt{cluster\_dilate\_radius=6}). We also clean these pixels with a 2D interpolation with kernel size of 7.
    \item Finally, we replace any remaining \texttt{NaN} values with 0.
\end{enumerate}

For a handful of targets, we used one additional custom pixel flagging step, manually identifying pixels that survive the cleaning routine above. We highlight that the specific steps and parameters used for pixel cleaning are likely dataset- and pipeline-dependent. 

We do PSF re-centering by performing a phase cross-correlation to the data with a \texttt{STPSF} model created with the spectral type of the each target. For this, we used \texttt{STPSF} version 2.1.1, using specifically the version in the \texttt{develop} branch (GitHub commit hash: 1abf9bf3). This specific version of \texttt{STPSF} includes updates to the optical aberrations in coronagraphic modes of JWST NIRCam, and likely results in more accurate centering estimates. We shift the data using a Fourier shifting routine. 

While all the observations can be re-centered independently in principle, it was found that the alignment was better when the science target (the one that will be PSF-subtracted) was re-centered individually and the targets in the PSF library were re-centered relative to the science target. For optimal results, then, re-centering (and therefore the following step of background subtraction) must be done on a per-science target basis before PSF subtraction. 

We perform a custom background subtraction routine that takes into account the varying throughput of the background level inside the coronagraph mask. See \cite{lawson2024} for details.

Finally, we mask the neutral density squares in our data to avoid biasing the PSF subtraction and our sensitivity calculation, and we pad the data with 40 pixels per side.

All of the steps described here are implemented in \texttt{spaceKLIP}.

\section{Pixel masking and replacement}
\label{s:app_masking}

In Section \ref{s:selection} we discuss our masking procedure to clean reference observations from off-axis astrophysical contaminants. In summary, we replace the pixels within a circular region enclosing the contaminant with the stellar flux-weighted pixel values of a template PSF. The template PSFs are the pixel-wise median of sets of observations free of contaminants. We present the details of these sets of observations in Table \ref{tab:forpsftemplate}.

Importantly, the observations in Table \ref{tab:forpsftemplate} are all contaminant-free, and thus they have not been masked. The total number of integrations used for each PSF template 1 through 10 is 18, 91, 10, 20, 20, 66, 29, 86, 14, 46. For other observations that were masked using these templates, the total number of masks across all observations in the F444W filter is 349, and 164 in the F200W filter.

The main goal of this masking and replacement process is minimizing contamination of our PSF-subtracted images, both by reducing biases in the KLIP basis calculation (by removing significant sources of signal that do not belong to the coronagraphic stellar PSF, and by trying to maintain stellar PSF structure if the masked region is close to the star), and by reducing the changes of false positive injections (by removing off-axis sources that could be injected into the PSF-subtracted images with positive flux, masquerading as planet candidate detections). As a result, this pixel replacement strategy is not necessarily expected to maintain local pixel statistics.

\begin{figure*}[h!]
    \centering
    \includegraphics[width=\linewidth]{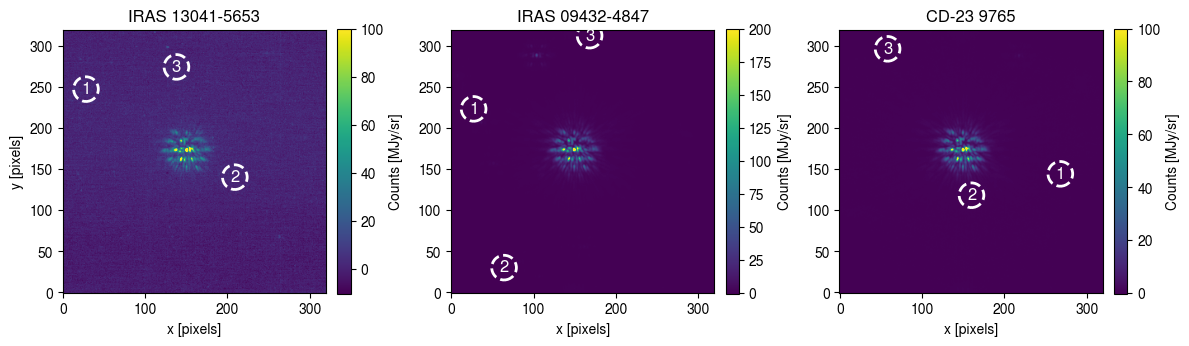}
    \includegraphics[width=\linewidth]{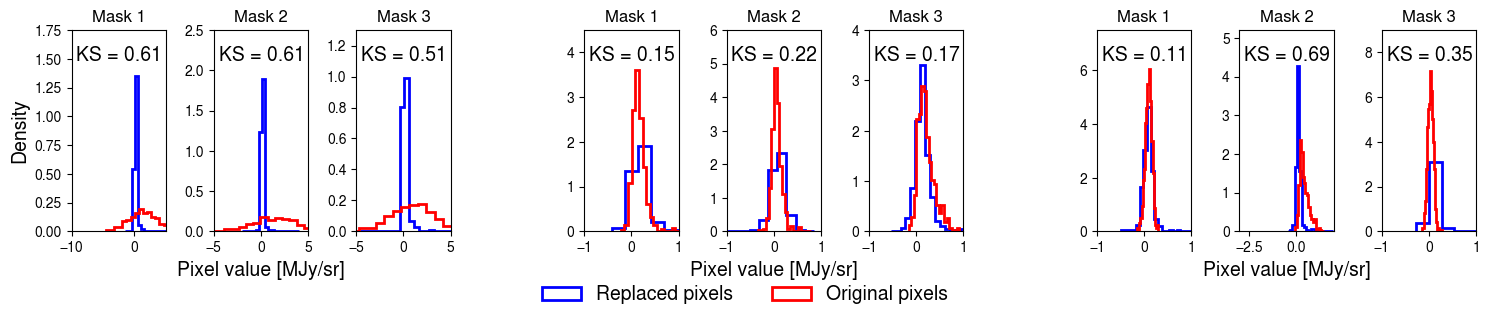}
    \includegraphics[width=\linewidth]{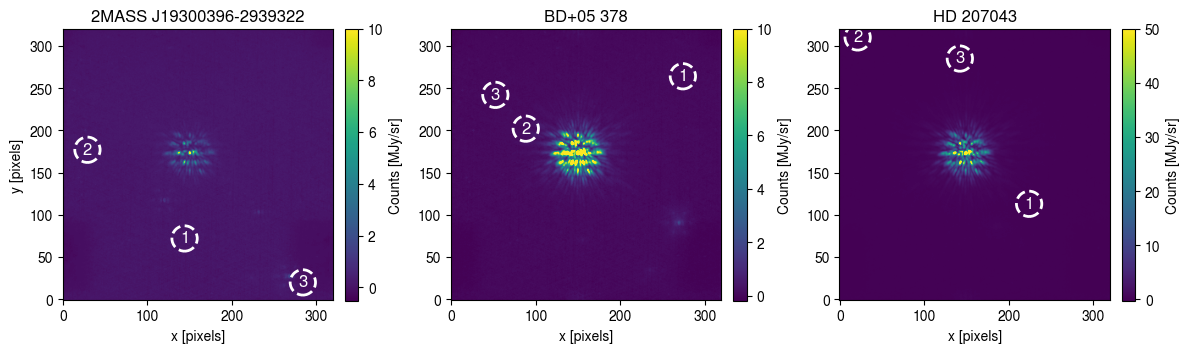}
    \includegraphics[width=\linewidth]{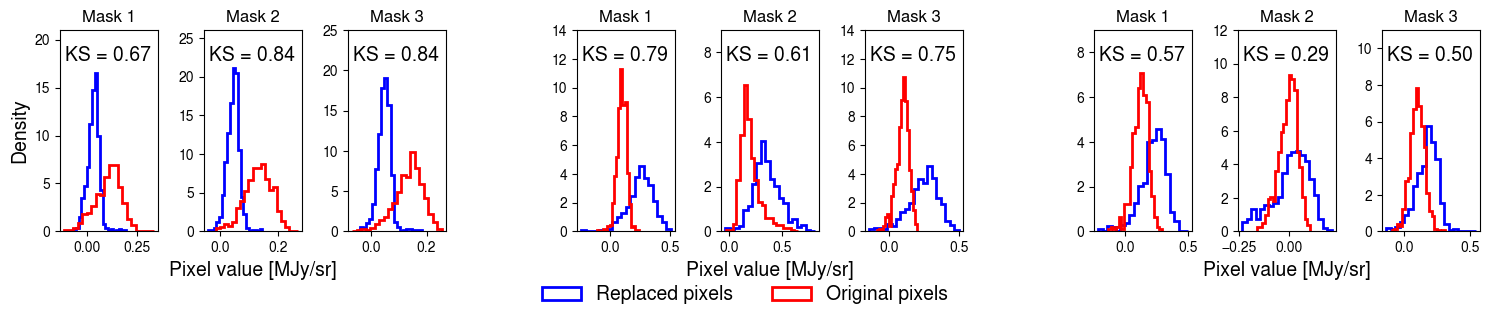}
    \caption{Local pixel statistics before and after applying our masking and replacement strategy. We show six random observations from our library, each of which has been masked on three random patches (radius of 15 pixels) with no significant astrophysical contaminants. Images are in the detector orientation (not North up). Below each image we show three histograms of pixel brightness (one per patch, as labeled) before and after pixel replacement. In each histogram panel, we show the Kolmogorov-Smirnov (KS) statistic comparing the pixel distribution before and after masking. Pixel statistics are not necessarily conserved before and after our masking process.}
    \label{fig:pixelstats}
\end{figure*}

To show the effect on the pixel statistics, we employ the same masking and replacement strategy on regions of the field of view free of contaminant sources, and see how the original pixels compare with the replaced pixels. We show this for six random observations from our library in Figure \ref{fig:pixelstats}. In each image, we mask and replace three random circular patches with radius of 15 pixels that are originally free of significant contaminants (each patch contains $\sim700$ individual pixels). We see that pixel statistics are not necessarily conserved before and after our masking process. This can be seen from the pixel brightness distributions themselves and from the Kolmogorov-Smirnov (KS) statistic (KS varies from 0.11 to 0.84). This is expected given the diversity in background SNRs and stellar brightness. For instance, for the Top Left panel of Figure \ref{fig:pixelstats}, our replaced pixels show a significantly narrower dispersion than the original pixels, since the original observation has relatively low background SNR. Likely the opposite occurs in the Bottom Center panel of the figure, where the observation being masked has higher SNR than the images used to construct the median PSF model. If one wished to maintain pixel statistics, particularly in the background-limited regime, a more complex masking strategy could be used. This could involve assessing the level of background SNR compared to the SNR of the stellar PSF, and substituting the masked pixels with pixel values that are randomly distributed following the original pixel distribution.

Nevertheless, our simple masking and replacement strategy is sufficient for our purposes in this work. In most cases the effects of altering these pixel statistics are small compared to contamination by astrophysical sources, which is the main goal of the pixel replacement process. The masked regions have generally small mean and dispersions, and importantly, their sharp circular shapes are unlikely to be mistaken by planet candidates even if they are propagated to the PSF-subtracted image (see TWA 23 in Figure \ref{fig:twa23} for an example).

Other than the impact on pixel statistics, a different question is whether the masking strategy used here can impact the achieved contrast by biasing the calculation of the principal components during KLIP. To assess this, we carry out a sample PSF subtraction where we randomly mask clean regions in our reference images. In every image of our PSF library, we randomly mask three circular patches with radius of 15 pixels, and we replace them using our pixel replacement strategy. This number of masks and their size is comparable to the masks applied to our real library when removing small, sparse astrophysical contaminants, but our random masking will mostly replace clean regions of the images free of contaminants (just like Figure \ref{fig:pixelstats}). We compare the contrast limits from using this library with extra masks to the contrast limits achieved with our real PSF library. We also show the contrast limits when none of the sources are masked, for completeness.

\begin{figure}
    \centering
    \includegraphics[width=\linewidth]{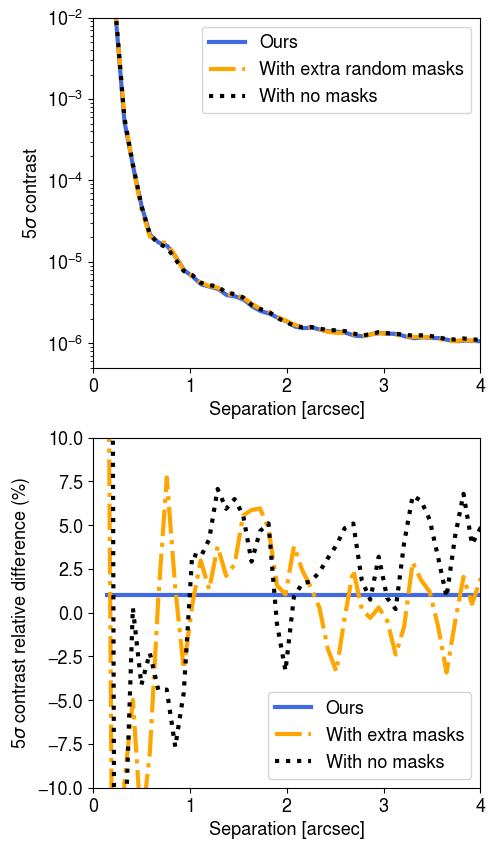}
    \caption{Effect of our source masking on the contrast limits for AP Col, as an illustrative example. Top: Contrast limits when using our library with masked sources, without masking any sources, and with masked sources and random masked clean regions in the image. Bottom: Relative contrast differences to our full library.  All KLIP reductions are done with the original PSF library from \cite{bogat2025probing} and with our full library. We see that our pixel masking strategy, when applied to clean regions in the image or to off-axis contaminants, has a negligible impact on the final contrast, as its main purpose is to avoid spurious injections (see text). All curves differ by less than $\sim5\%$ at most separations.}
    \label{fig:contrastrandmask}
\end{figure}

We show the result of this test in Figure \ref{fig:contrastrandmask} for AP Col. We see that the contrast limits from our real library are nearly identical to the limits from using our library with extra masks. This shows that, even when replacing clean pixels in all reference images, a moderate number of small masks has little to no negative effect on sensitivity. This highlights the fact that our masking strategy accomplishes its goal of reducing the number of contaminants and false positives injected by the PSF library, with little to no negative impact on the overall sensitivity. 

On the other hand, drastically increasing the number of masks or the size of each mask is likely to negatively affect the subtraction. This is clear in the hypothetical extreme case where all pixels in all reference images are replaced. It is for this reason that we only include in our PSF library the observations that have at most a small number of small contaminants, so they can be effectively masked without replacing a significant portion of the images.


\startlongtable
\begin{deluxetable*}{cccccccc}
\tabletypesize{\small}
\tablewidth{\textwidth}
\tablecaption{Observations used for each PSF template \label{tab:forpsftemplate}}
\tablehead{
\colhead{Target name} & \colhead{Filter}& \colhead{Date}&\colhead{Readout pattern} & \colhead{$N_i$/$N_g$/$N_f$} & 
\colhead{$t_{\text{int}}$ (s)} & \colhead{$t_{\text{exp}}$ (s)} & 
\colhead{PSF template}
}
\startdata
IRAS 13041-5653 (4) & F444W & 2024-06-06 & RAPID & 2/5/1 & 5 & 10 & 1 \\
IRAS 13041-5653 (1) & F444W & 2024-06-06 & RAPID & 2/5/1 & 5 & 10 & 1 \\
IRAS 13041-5653 (7) & F444W & 2024-06-06 & RAPID & 2/5/1 & 5 & 10 & 1 \\
IRAS 13041-5653 (5) & F444W & 2024-06-06 & RAPID & 2/5/1 & 5 & 10 & 1 \\
IRAS 13041-5653 (9) & F444W & 2024-06-06 & RAPID & 2/5/1 & 5 & 10 & 1 \\
IRAS 13041-5653 (8) & F444W & 2024-06-06 & RAPID & 2/5/1 & 5 & 10 & 1 \\
V* Y Crt (3) & F444W & 2024-06-15 & SHALLOW4 & 2/4/4 & 20 & 40 & 1 \\
IRAS 13041-5653 (3) & F444W & 2024-06-06 & RAPID & 2/5/1 & 5 & 10 & 1 \\
IRAS 13041-5653 (2) & F444W & 2024-06-06 & RAPID & 2/5/1 & 5 & 10 & 1 \\
CD-23 9765 (5) & F200W & 2024-02-04 & DEEP8 & 2/12/8 & 243 & 487 & 2 \\
HD 101581 (6) & F200W & 2024-06-16 & DEEP8 & 2/6/8 & 115 & 230 & 2 \\
HD 101581 (2) & F200W & 2024-06-16 & DEEP8 & 2/6/8 & 115 & 230 & 2 \\
HD 101581 (5) & F200W & 2024-06-16 & DEEP8 & 2/6/8 & 115 & 230 & 2 \\
CD-23 9765 (6) & F200W & 2024-06-15 & DEEP8 & 2/12/8 & 243 & 487 & 2 \\
V* Y Crt (6) & F200W & 2024-06-15 & SHALLOW4 & 2/4/4 & 20 & 40 & 2 \\
HD 101581 (3) & F200W & 2024-06-16 & DEEP8 & 2/6/8 & 115 & 230 & 2 \\
CD-23 9765 (3) & F200W & 2024-02-04 & DEEP8 & 2/12/8 & 243 & 487 & 2 \\
TWA 36 & F200W & 2024-06-15 & DEEP8 & 5/18/8 & 372 & 1860 & 2 \\
HD 101581 (1) & F200W & 2024-06-16 & DEEP8 & 2/6/8 & 115 & 230 & 2 \\
TWA 35 & F200W & 2024-06-15 & DEEP8 & 5/18/8 & 372 & 1860 & 2 \\
V* Y Crt (5) & F200W & 2024-06-15 & SHALLOW4 & 2/4/4 & 20 & 40 & 2 \\
CD-23 9765 (1) & F200W & 2024-02-04 & DEEP8 & 2/12/8 & 243 & 487 & 2 \\
CD-23 9765 (5) & F200W & 2024-06-15 & DEEP8 & 2/12/8 & 243 & 487 & 2 \\
V* Y Crt (4) & F200W & 2024-06-15 & SHALLOW4 & 2/4/4 & 20 & 40 & 2 \\
CD-23 9765 (2) & F200W & 2024-02-04 & DEEP8 & 2/12/8 & 243 & 487 & 2 \\
CD-23 9765 (1) & F200W & 2024-06-15 & DEEP8 & 2/12/8 & 243 & 487 & 2 \\
V* Y Crt (9) & F200W & 2024-06-15 & SHALLOW4 & 2/4/4 & 20 & 40 & 2 \\
V* Y Crt (2) & F200W & 2024-06-15 & SHALLOW4 & 2/4/4 & 20 & 40 & 2 \\
TWA 33 & F200W & 2024-06-15 & DEEP8 & 5/18/8 & 372 & 1860 & 2 \\
CD-23 9765 (4) & F200W & 2024-02-04 & DEEP8 & 2/12/8 & 243 & 487 & 2 \\
V* Y Crt (1) & F200W & 2024-06-15 & SHALLOW4 & 2/4/4 & 20 & 40 & 2 \\
HD 101581 (8) & F200W & 2024-06-16 & DEEP8 & 2/6/8 & 115 & 230 & 2 \\
V* Y Crt (8) & F200W & 2024-06-15 & SHALLOW4 & 2/4/4 & 20 & 40 & 2 \\
CD-23 9765 (3) & F200W & 2024-06-15 & DEEP8 & 2/12/8 & 243 & 487 & 2 \\
TWA 45 & F200W & 2024-06-15 & DEEP8 & 5/18/8 & 372 & 1860 & 2 \\
V* Y Crt (7) & F200W & 2024-06-15 & SHALLOW4 & 2/4/4 & 20 & 40 & 2 \\
* 10 Crt & F200W & 2024-06-15 & DEEP8 & 9/10/8 & 200 & 1808 & 2 \\
CD-23 9765 (2) & F200W & 2024-06-15 & DEEP8 & 2/12/8 & 243 & 487 & 2 \\
CD-23 9765 (9) & F200W & 2024-06-16 & DEEP8 & 2/12/8 & 243 & 487 & 2 \\
V* Y Crt (3) & F200W & 2024-06-15 & SHALLOW4 & 2/4/4 & 20 & 40 & 2 \\
HD 101581 (7) & F200W & 2024-06-16 & DEEP8 & 2/6/8 & 115 & 230 & 2 \\
CD-23 9765 (4) & F200W & 2024-06-15 & DEEP8 & 2/12/8 & 243 & 487 & 2 \\
HD 101581 (9) & F200W & 2024-06-16 & DEEP8 & 2/6/8 & 115 & 230 & 2 \\
HD 101581 (4) & F200W & 2024-06-16 & DEEP8 & 2/6/8 & 115 & 230 & 2 \\
CD-23 9765 (7) & F200W & 2024-06-15 & DEEP8 & 2/12/8 & 243 & 487 & 2 \\
RX J0019.7+1951 & F444W & 2024-11-19 & DEEP8 & 5/18/8 & 372 & 1860 & 3 \\
SCR J0017-6645 & F444W & 2024-10-02 & DEEP8 & 5/18/8 & 372 & 1860 & 3 \\
PM J21214-6655 & F200W & 2024-09-17 & DEEP8 & 5/18/8 & 372 & 1860 & 4 \\
SCR J0017-6645 & F200W & 2024-10-02 & DEEP8 & 5/18/8 & 372 & 1860 & 4 \\
RX J0613.2-2742 & F200W & 2024-10-03 & DEEP8 & 5/18/8 & 372 & 1860 & 4 \\
PM J22088+1144 & F200W & 2024-11-19 & DEEP8 & 5/18/8 & 372 & 1860 & 4 \\
MCC 124 & F444W & 2025-05-15 & DEEP8 & 7/18/8 & 372 & 2604 & 5 \\
V* WW PsA & F444W & 2025-05-15 & DEEP8 & 7/18/8 & 372 & 2604 & 5 \\
BD-13 6424 & F444W & 2025-07-14 & DEEP8 & 6/18/8 & 372 & 2232 & 5 \\
Barta 161 12 & F200W & 2025-07-15 & DEEP8 & 7/18/8 & 372 & 2604 & 6 \\
GR* 9 & F200W & 2025-07-15 & DEEP8 & 6/18/8 & 372 & 2232 & 6 \\
BD-13 6424 & F200W & 2025-07-14 & DEEP8 & 6/18/8 & 372 & 2232 & 6 \\
PM J00502+0837 & F200W & 2025-07-14 & DEEP8 & 6/18/8 & 372 & 2232 & 6 \\
GJ 2006 A & F200W & 2025-07-14 & DEEP8 & 7/18/8 & 372 & 2604 & 6 \\
V* WW PsA & F200W & 2025-05-15 & DEEP8 & 7/18/8 & 372 & 2604 & 6 \\
HD 207043 & F200W & 2025-09-20 & DEEP8 & 6/18/8 & 372 & 2232 & 6 \\
SIPS J2335-3401 & F200W & 2025-07-15 & DEEP8 & 7/18/8 & 372 & 2604 & 6 \\
MCC 124 & F200W & 2025-05-15 & DEEP8 & 7/18/8 & 372 & 2604 & 6 \\
SIPS J1809-7613 & F200W & 2025-08-20 & DEEP8 & 7/18/8 & 372 & 2604 & 6 \\
2MASS J01162926+4256219 & F444W & 2024-08-29 & DEEP8 & 9/10/8 & 200 & 1808 & 7 \\
2MASS J05203803-3945179 & F444W & 2024-11-26 & DEEP8 & 5/18/8 & 372 & 1860 & 7 \\
2MASS J23301372+3142210 & F444W & 2024-11-28 & DEEP8 & 5/18/8 & 372 & 1860 & 7 \\
2MASS J00135300-7441178 & F444W & 2024-09-05 & DEEP8 & 5/18/8 & 372 & 1860 & 7 \\
2MASS J00345120-6154583 & F444W & 2024-10-16 & DEEP8 & 5/18/8 & 372 & 1860 & 7 \\
2MASS J04152578+0611588 & F200W & 2024-10-06 & DEEP8 & 9/10/8 & 200 & 1808 & 8 \\
2MASS J05203803-3945179 & F200W & 2024-11-26 & DEEP8 & 5/18/8 & 372 & 1860 & 8 \\
2MASS J04213348-2543091 & F200W & 2024-09-02 & DEEP8 & 5/18/8 & 372 & 1860 & 8 \\
2MASS J04152879+0611128 & F200W & 2024-10-06 & DEEP8 & 9/10/8 & 200 & 1808 & 8 \\
2MASS J03304909-4555573 & F200W & 2024-08-29 & DEEP8 & 5/18/8 & 372 & 1860 & 8 \\
2MASS J21310171+2320074 & F200W & 2024-11-17 & DEEP8 & 5/18/8 & 372 & 1860 & 8 \\
2MASS J01162926+4256219 & F200W & 2024-08-29 & DEEP8 & 9/10/8 & 200 & 1808 & 8 \\
2MASS J00135300-7441178 & F200W & 2024-09-05 & DEEP8 & 5/18/8 & 372 & 1860 & 8 \\
2MASS J23030499+2055068 & F200W & 2024-12-13 & DEEP8 & 9/10/8 & 200 & 1808 & 8 \\
2MASS J00345120-6154583 & F200W & 2024-10-16 & DEEP8 & 5/18/8 & 372 & 1860 & 8 \\
2MASS J04384393-2702018 & F200W & 2024-12-25 & DEEP8 & 5/18/8 & 372 & 1860 & 8 \\
2MASS J03190864-3507002 & F200W & 2024-09-23 & DEEP8 & 5/18/8 & 372 & 1860 & 8 \\
2MASS J01574896-2154052 & F200W & 2024-08-23 & DEEP8 & 5/18/8 & 372 & 1860 & 8 \\
2MASS J00251465-6130483 & F200W & 2024-08-27 & DEEP8 & 5/18/8 & 372 & 1860 & 8 \\
2MASS J07330059+3701475 & F444W & 2025-04-01 & DEEP8 & 9/10/8 & 200 & 1808 & 9 \\
2MASS J15353016+3612347 & F444W & 2025-04-13 & DEEP8 & 5/18/8 & 372 & 1860 & 9 \\
2MASS J21130526-1729126 & F200W & 2025-06-15 & DEEP8 & 5/18/8 & 372 & 1860 & 10 \\
2MASS J12122084+5429086 & F200W & 2025-04-02 & DEEP8 & 5/18/8 & 372 & 1860 & 10 \\
2MASS J15353016+3612347 & F200W & 2025-04-13 & DEEP8 & 5/18/8 & 372 & 1860 & 10 \\
2MASS J07234358+2024588 & F200W & 2025-03-29 & DEEP8 & 5/18/8 & 372 & 1860 & 10 \\
2MASS J07330059+3701475 & F200W & 2025-04-01 & DEEP8 & 9/10/8 & 200 & 1808 & 10 \\
2MASS J06064047+1532317 & F200W & 2025-03-17 & DEEP8 & 9/10/8 & 200 & 1808 & 10 \\
2MASS J12075091+1856558 & F200W & 2025-06-17 & DEEP8 & 5/18/8 & 372 & 1860 & 10 \\
\enddata
\tablecomments{The PSF template number refers to different sets of observations grouped by similarity, as described in Section \ref{s:selection}. PSF templates 1 and 2 are for program GO 4050, templates 3 to 6 for GO 5835, and 7 to 10 for SURVEY 6005. Observations in F444W and in F200W are counted separately as parts of different sets. The PSF template for each set is computed with a pixel-wise median across all integrations of all observations in a given set. Target names with parenthesis correspond to dedicated reference stars in GO 4050, with the number in parenthesis indicated the dither position. $N_i$, $N_g$ and $N_f$ refers to number of integrations, number of groups and number of frames, respectively.}
\end{deluxetable*}

\section{Algorithm throughput calibration}
\label{s:app_algothroughput}

\begin{figure*}
    \centering
    \includegraphics[width=0.48\linewidth]{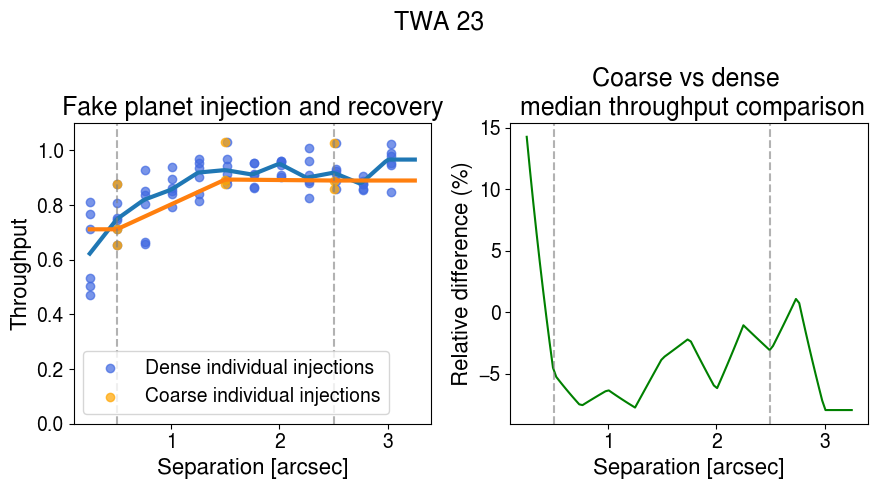}
    \includegraphics[width=0.48\linewidth]{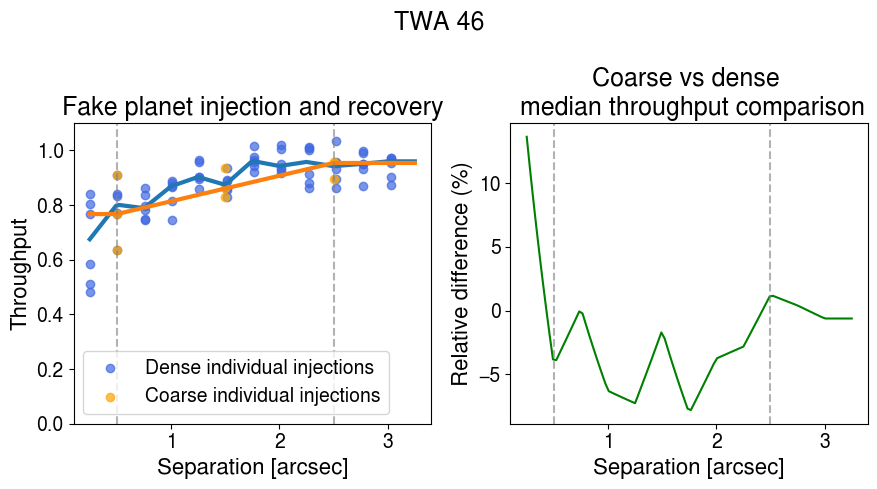}
    \caption{Algorithm throughput calibration comparison between two choices of fake planet injections, in the context of our analysis in Section \ref{s:results_grids}. We show two targets from our library, TWA 23 and TWA 46, both reduced with our full PSF reference library using 3 annuli, all KL modes and all \texttt{maxnumbasis}. We compare a ``coarse" and a ``dense" set of injections, where the dense strategy uses more separations and more position angles per separation (see text). Solid lines interpolate the median throughput per separation, computed across position angles. Our analysis in Section \ref{s:results_grids} uses the coarse strategy to save computation time. We see that at the separations where we evaluate sensitivity for Figures \ref{fig:calcon05} and \ref{fig:calcon3} (dashed vertical lines), both strategies agree at the $\sim5\%$ level or better.}
    \label{fig:throughputs_comparesampling}
\end{figure*}

In Section \ref{s:results_grids}, we compare the injection-recovery-calibrated contrast limits across targets in our library to discern preferred KLIP hyper-parameters (Figures \ref{fig:calcon05} and \ref{fig:calcon3}). Due to the large number of datasets processed, we opted to calibrate our contrast using only three separations (0.5, 1.5, 2.5 arcsec) and three position angles per separation (0, 120, 240 deg). Here we perform a brief validation of this choice on two datasets, which we show in Figure \ref{fig:throughputs_comparesampling}. For this, we compare our ``coarse" planet injection strategy with a ``dense" strategy, where we significantly increase the number of separations and position angles, and compare the resulting throughput corrections between both approaches. For the ``dense" strategy, we use separations from 0.25 to 3.0 arcsec in steps of 0.25 arcsec, and at each separation we use six position angles (equally spaced azimuthally in steps of 60 deg). We do this test on TWA 23 and TWA 46, both targets part of our PSF library.

We see that at most separation injections, the dense and coarse injections agree within $\sim5\%$. The only significant exception is at the innermost separation of 0.25 arcsec, where assuming the same throughput as 0.5 arcsec for the coarse strategy yields errors $>10\%$. For the separations probed in Figures \ref{fig:calcon05} and \ref{fig:calcon3}, both of which require no interpolation since planets were injected at those separations, we see good agreement between the coarse and dense throughput calibrations. This supports our choice to adopt the coarse strategy in Section \ref{s:results_grids}, since our results are not strongly dependent on the number of injected fake planets.

\begin{figure*}
    \centering
    \includegraphics[width=\linewidth]{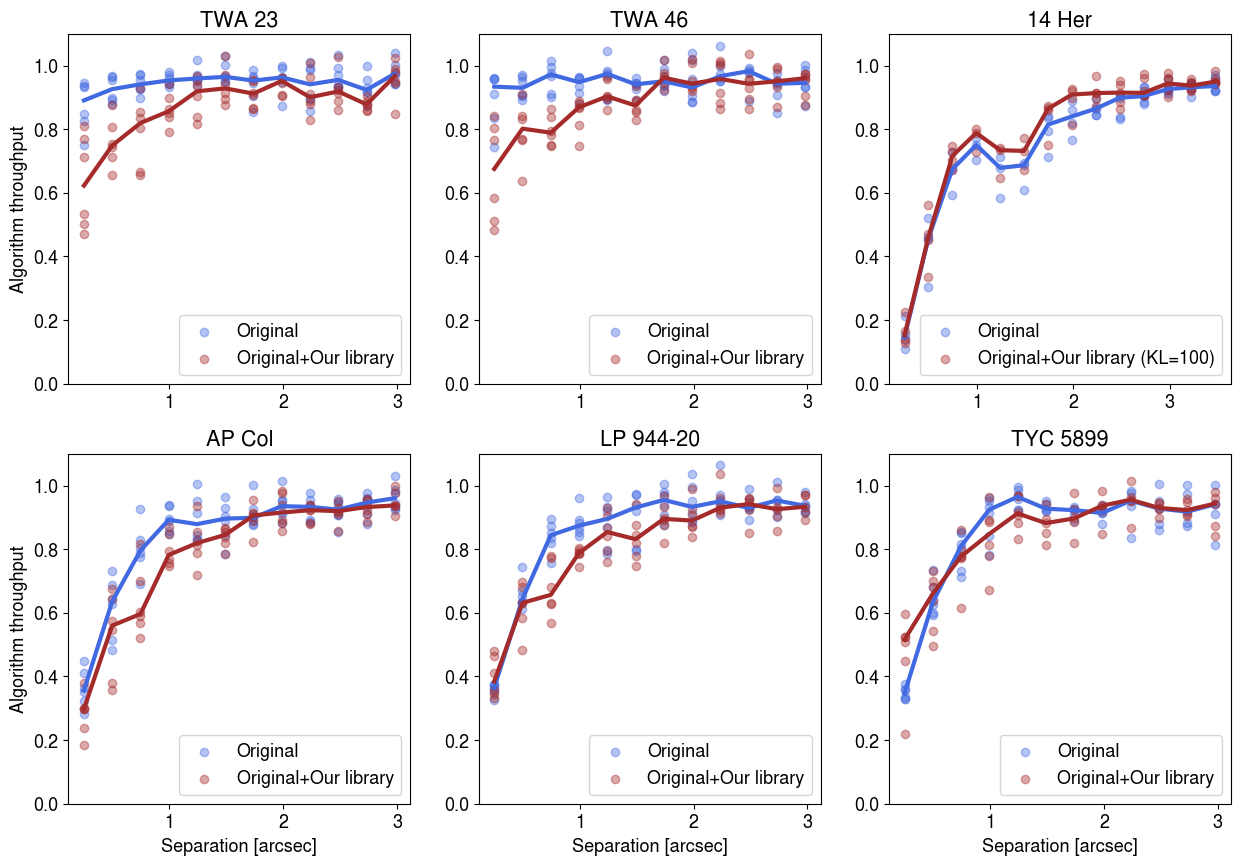}
    \caption{Algorithm throughput for all targets shown in Figures \ref{fig:twa23} and \ref{fig:ellsexamples}. Shown are the individual injections and recoveries of fake planets, and the lines represent the median throughput across position angles per separation. The median throughput is the correction applied to the raw contrast curves. All ``Original" curves correspond to the KLIP parameters described in the main text. All ``Original+Our Library" correspond to the panels in the second row in Figure \ref{fig:twa23} and \ref{fig:ellsexamples}.}
    \label{fig:allthroughputs}
\end{figure*}

Figure \ref{fig:allthroughputs} shows the throughput correction for the nominal contrast curves used in Figures \ref{fig:twa23} and \ref{fig:ellsexamples}. In most cases, we see that using our big library leads to slightly lower planet throughput. This is particularly visible in TWA 23 and TWA 46 where there is no significant self-subtraction due to ADI (since these two observations are only processed with RDI). In those two targets we can see a substantial decrease in throughput, particularly at close separations. This is expected given the larger available complexity in building the PSF model with a large library and using all KL modes. We note that in 14 Her our choice to use only 100 KL modes helps preserve more planet throughput than using all KL modes with the dedicated reference star. In datasets with ADI, the difference in throughput with a large library at very close separations (within $\sim0.5$ arcsec) may be small compared to the expected low throughput resulting from ADI self-subtraction (within uncertainties from throughput dispersion).  We also note that at closer separations there is larger fractional variability in the throughput: this partly explains the larger error bars seen in the contrast improvement plots at close angular separations (bottom panels in Figures \ref{fig:twa23} and \ref{fig:ellsexamples}).

\bibliography{sample701}{}
\bibliographystyle{aasjournalv7}



\end{document}